\documentclass[onefignum,onetabnum]{siamsials250905}

\usepackage{lipsum}
\usepackage{amsfonts}
\usepackage{graphicx}
\usepackage{epstopdf}
\usepackage{algorithmic}
\ifpdf
  \DeclareGraphicsExtensions{.eps,.pdf,.png,.jpg}
\else
  \DeclareGraphicsExtensions{.eps}
\fi

\newsiamremark{remark}{Remark}
\newsiamremark{hypothesis}{Hypothesis}
\crefname{hypothesis}{Hypothesis}{Hypotheses}
\newsiamthm{claim}{Claim}

\headers{Inferring Relative Consequences of Mechanical Ventilation}{Albers, Bennett, {d}e Wiljes, Hripcsak, Smith, Sottile, Stroh}

  \title{Inferring Relative Consequences of Mechanical Ventilation from Observational Data Using Game-Based Comparisons\thanks{This work is supported by National Heart Lung and Blood Institute awards 5R01HL151630 ``Predicting and Preventing Ventilator-Induced Lung Injury'',  K23HL145011 ``The Detection, Quantification, and Management of Ventilator Dyssynchrony'', and K24HL168225 ``Mentoring and Patient-Oriented Research in Clinical Informatics and Data Science" as well as National Library of Medicine award R01LM006910 ``Discovering and Applying Knowledge in Clinical Databases''. The research of JdW has been partially funded by the Deutsche Forschungsgemeinschaft (DFG)- Project-ID 318763901 - SFB1294. Additionally JdW gratefully acknowledges funding by the Carl-Zeiss foundation within the project KI-MSO-O.}
 }

\author{
David J. Albers\thanks{Department of Biomedical Informatics, University of Colorado Anschutz, Aurora, CO  USA; Department of Biomedical Informatics, Columbia University, New York, NY  USA}
\and 
Tellen D. Bennett\thanks{Department of Biomedical Informatics, University of Colorado Anschutz, Aurora, CO  USA; Pediatrics and Critical Care, Children's Hospital Colorado, Aurora, CO  USA}
 \and Jana de Wiljes\thanks{Department of Mathematics and Natural Sciences, Technische Universit\"{a}t Ilmenau, Ilmenau, DE; School of Engineering Sciences, Department of Computational Engineering, LUT University, Lappeenranta, FI}
\and 
George Hripcsak\thanks{Department of Biomedical Informatics, Columbia University, New York, NY  USA}\and 
Bradford J. Smith\thanks{Department of Biomedical Engineering, University of Colorado Denver \textbar Anschutz Medical Campus, Aurora, CO USA}
\and  Peter D. Sottile\thanks{Pulmonary Sciences and Critical Care Medicine, University of Colorado Anschutz, Aurora, CO USA}
\and J.N. Stroh\thanks{Department of Biomedical Informatics, University of Colorado Anschutz, Aurora, CO  USA}
}

\usepackage{amsopn}

\makeatletter
\newcommand*{\addFileDependency}[1]{
  \typeout{(#1)}
  \@addtofilelist{#1}
  \IfFileExists{#1}{}{\typeout{No file #1.}}
}
\makeatother

\ifpdf
\hypersetup{
	pdftitle={Inferring Relative Consequences of Mechanical Ventilation from Observational Data Using Game-Based Comparisons},
	pdfauthor={D.J. Albers, T.D. Bennett, J. de Wiljes, G. Hripcsak, B.J. Smith, P.D. Sottile, J.N. Stroh}
}
\fi

\usepackage{xspace}
\usepackage{tikz-cd} 
		\usetikzlibrary{graphs,decorations.pathmorphing,decorations.markings,shapes.geometric,fit}

\begin{document}
	\maketitle
	
\begin{abstract}
Identifying the effects of mechanical ventilation (MV) policies and protocols in critical care requires analyzing data from heterogeneous patient--ventilator systems within the context of the clinical decision-making environment.
The coupled, multiscale interactions among these system components generate a high-dimensional state space that remains sparsely sampled despite extensive clinical data collection.
Analysis of existing data is nevertheless essential for understanding current respiratory management practices and generating testable hypotheses about improvement. 
The scale and complexity of available data motivate the use of reinforcement learning (RL) to explore data-consistent counterfactual trajectories.
However, formulating RL in practical applications requires a spatiotemporally dependent reward process that defines state-to-consequence relationships, their context dependence, and the delays over which consequences emerge. 
These poorly understood elements are not known \emph{a priori} and must instead be hypothesized and inferred from data.
To that end, categorized observed states are contrasted according to their relative consequences by solving a game-based inverse problem that identifies a comparison model required for downstream probabilistic and stochastic methods such as reinforcement learning.
This development is one step along the pathway toward MV optimization and personalization. 
The inverted-game inference is analytically validated on synthetic data to reveal potential caveats before proceeding to real-world ICU data applications that expose complexities of the data-generating process.
Clinical data applications revealed that both breath-type consequences and their relative ordering are inherently context- and time-dependent, varying across patient subgroups, time, and  definitions of MV consequences including the effect timescale.
The discussion includes potential developments toward a state transition model for simulating the effects of MV management actions using empirical data and game-inferred comparisons.
\end{abstract}

\begin{relevance}
This work, including real-data applications, investigates the lung health-related consequences of breath behaviors that occur under mechanical ventilation, a crucial element of human critical care medicine.
It contributes to a broader effort to understand and improve the clinical decisions made by medical professionals about the joint management of patients and mechanical ventilator therapy.
\end{relevance}

\begin{mathcontent}
This work uses games to formulate comparison of windowed event occurrence distributions on the basis of their cost or consequences. 
The associated inverse problem is posed and solved to identify consequence differences among event classes, which are needed for downstream reinforcement learning applications.
\end{mathcontent}
	


\section{Introduction} 
\label{sec:intro}
Mechanical ventilation (MV) is a key life-sustaining patient therapy based on a rule-based apparatus that exposes patients to iatrogenic risks of insult and injury via ventilator-induced lung injury (VILI). 
VILI and MV contribute to or exacerbate acute respiratory distress syndrome (ARDS\cite{ranieri2012acute}), a condition associated with over 20\% of all MV encounters and an in-hospital mortality rate of 30--50\% \cite{bellani2016epidemiology,cochi2016mortality}. 
MV management poses a complex clinical problem, requiring healthcare providers to minimize VILI while maintaining adequate gas exchange and respiratory function \cite{pham2017mechanical}.
However, the complexity of the patient--ventilator system, the clinical decision process, and their mutual interactions impedes the systematic study needed to improve MV management.

The patient--ventilator system is a hybrid interactive biomechanical system that limits rule-based modelability. 
Additionally, ventilator management is guided by protocols with care provider deviations based on wider treatment objectives.
Both components involve interaction between rule-based systems (ventilators, protocols) and free-acting agents (patients, doctors) that are coupled through patient-doctor interaction (non-MV care). 
Patient outcomes emerge from reactive and proactive actions taken regarding MV and other entangled patient care processes.
Any realistic method linking patient outcomes to MV actions (management decisions) must then consider external data (\textit{e.g.}, sedation and other medications, posture, and other interventions) beyond ventilator settings.
This work uses formal games to compare complex system behavior categories on the basis of intermediate clinical outcomes, and advances the action-to-consequences problem by identifying the relative costs of system states by solving the associated inverse game problem.
The results permit more practical application of reinforcement learning (RL) that motivates the approach.

\subsection{Actions-to-consequences in the data-generating system}
The action-to-consequence linkage requires a broader view of the observable system because the behaviors and clinical consequences of mechanical ventilation emerge from interactions among the patient, ventilator, and care processes rather than from any component in isolation. 
Each MV patient is a cyborg \cite{clynes1960cyborgs,madrigal2010theman} formed by its biomechanical patient--ventilator system (PVS; the lung--ventilator system in previous work) \cite{stroh2023hypothesis,stroh2024identifying}. 
Care decisions about the cyborg (whether directed at the patient or ventilator) are co-predicated on patient status and history, disease, and clinical protocols along with constraints imposed by provider experience, institutional practice, and broad legal and logistical considerations. 
Rather than viewing patient--ventilator--care systems independently, this work treats protocol-guided clinical management as the primary agent and data generation system.
This way, clinical trajectories emerge from the coupled interactions of provider decisions,  patient physiology and idiosyncrasy, disease progression, and other factors framing emergent  consequences of ventilation.

\begin{figure} 
    \centering
    \includegraphics[width=.75\textwidth]{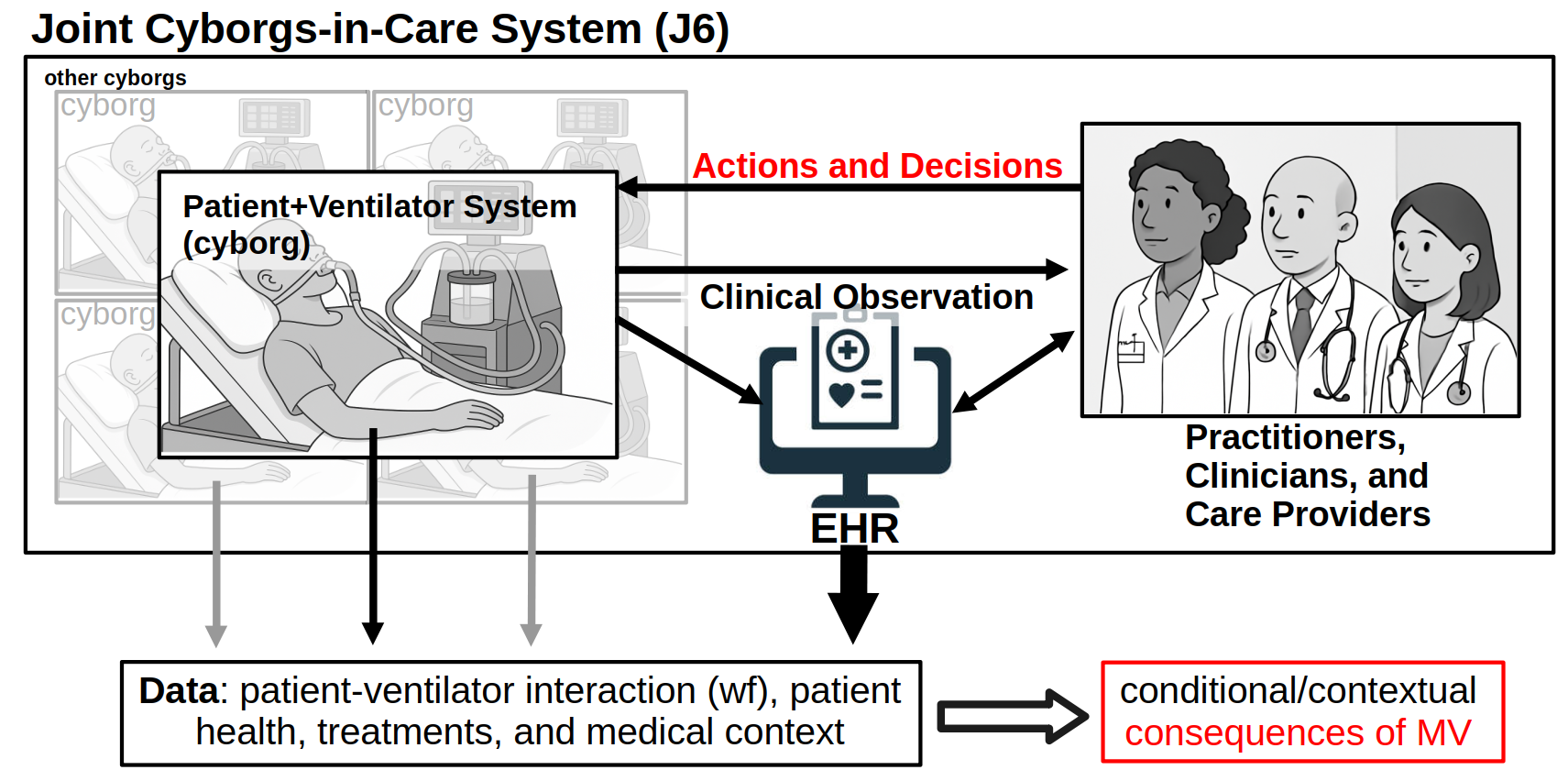}
    \caption{  
    Linking MV management to patient-health consequences that include VILI requires considering care decision impacts on the patient--ventilator cyborg.
    The system of interest is the joint cyborgs-in-care system (J6) depicted here, in which each joint patient+ventilator+care processes exists within a common care decision environment.    
    PVS cyborg behaviors are independent, but aspects of their evolution (ventilator settings, position, sedation, \textit{etc}.) are determined by common care practices and protocols. 
    Evaluation of these protocols' impact on observed PVS behaviors and patient outcome is the wider focus of this research. \textit{abbreviations}:  PVS=patient--ventilator system, MV=mechanical ventilation, wf=waveform
         }\label{fig:cz}
\end{figure}

The object of interest is the resulting \emph{joint cyborgs-in-care system} (JCICS, phonetically ``J-six" and referred to simply as J6), comprising the ensemble of patient trajectories generated under this coupled care process. 
Figure~\ref{fig:cz} depicts J6 and the pathway from care decisions within the broader clinical environment to their downstream consequences on the patient. 
Pressure ($p$), volume ($V$), and flow ($dV/dt$) waveforms capture the emergent patient--ventilator behaviors through which VILI and other downstream effects manifest. 
Because observed consequences emerge from the interaction of disease, patient physiology, and protocol-guided clinical care, whose underlying influences may be cooperative or competing and are not directly observable, J6 is treated as the observable system from which action-to-consequence relationships are to be inferred.

\subsection{Motivating challenges of the clinical objective}
Rather than modeling the latent processes governing J6 directly, RL instead operates on observed patient trajectories and their consequences, treating MV management as a time-dependent reward/penalty optimization problem \cite{SuttonBarto2018}.
The overall clinical objective translates to finding satisfactory MV management policies for actions (\textit{e.g.}, ventilator settings, sedation, positioning) that control patient--ventilator trajectories in a VILI-minimizing way.
Most previous RL applications address simplified decisions (binned vent settings or sedation), narrowly scoped problems (to wean or not) over very limited state spaces, and global outcome \cite{yu2020supervised,kondrup2023towards,liu2024reinforcement,safaei2024x}.
Recent works \cite{peine2021development,roggeveen2024reinforcement} have incorporated lung-specific status or VILI indicators as intermediate rewards but have yet to feature VILI minimization as an MV management objective.
Consistent with clinical decision making, the reinforcement learning reward process should compare multiple treatment options simultaneously \cite{lazo2012comparing}, producing context-dependent relative preferences rather than an absolute measure imposing a global rank over all options; however, no such comparison process is currently known.

This work introduces a game-theory inspired inverse framework for analyzing MV data to address two gaps in formulating a practical RL problem for MV management. 
By treating local breath behaviors as alternatives to be compared, the framework infers context-specific payoff matrices from observational data that quantify the relative consequences of breath behaviors under current care practices.
The output initiates differential penalties for RL in complex MV decision-making, but it first requires quantifying time-dependent health impacts and managing heterogeneity.
These challenges stem from the absence of local VILI biomarkers and the complexity of MV patient trajectories:
\paragraph{Quantifying MV consequences:} 
No definitive \textit{in vivo} biomarkers currently exist to quantify the short-term effects of VILI or MV, while global clinical outcomes (\textit{e.g.,} 90-day mortality, in-care mortality, mean days-since-last-ICU stay, or ventilator-free-days-in-28 \cite{yehya2019reappraisal}) cannot be reduced to incremental effects.
However, VILI could be proxied by common clinical observations (driving pressure, FiO\textsubscript{2}, and  P\textsubscript{a}O\textsubscript{2}:FiO\textsubscript{2} ratio (P:F); see \S\ref{ssec:consequences}) or by more targeted observation processes (\textit{e.g.}, tissue dissipation of ventilator energy\cite{gaver2025mechanical}). 

\paragraph{Heterogeneity:}    	
The myriad origins of J6 heterogeneity produce trajectories too diverse to analyze directly, generating heterogeneous treatment effects (HTEs) where identical ventilator decisions have different consequences depending on patient state. 
Care and ventilator conditions defining a J6 state may also require different and incompatible coordinate representations.
This work compartmentalizes heterogeneity through a two-stage partitioning of the J6 data: 
first, \emph{contexts} (\S\ref{ssec:contexts}) are defined from a priori hypotheses of state comparability; second, \emph{breath categories} (\S\ref{ssec:ptypes}) discretize the observations within each context based on feature similarity.

\subsection{Outline of this work:}
Section 2 describes the EHR and bedside records comprising J6 data, the comparison of J6 behaviors and MV consequences through a game formulated within a given context, and how payoff matrices for games are empirically identified (additional details in SI). 
Section 3 presents an internal verification to assess limitations of the method followed by the results of application using game-base inference on clinical J6 data. 
The final section discusses implications of the results and framing of a more general RL problem based on game results.

\section{Problem Formulation} \label{sec:meth}
Relating observed J6 behaviors to their local consequences involves mapping between observed data elements.
Few hypotheses exist for defining a manageable state space for high-dimensional, heterogeneous J6, and fewer for the functional form of the state-to-consequence mapping.
The basic problem is sketched here and formulated more fully in this section.
For now, let $\{\phi\}_{k=1}^K$ be elements of a discrete state space, $\pi_j$ be the state distribution over some $j$th time window, and $q_j\in\mathbb{R}$ be the associated consequences of $\pi_j$;
A na\"{i}ve regression approach, solving $[q_j] = \ell^\top[\pi_j]$, identifies consequences-per-state occurrence $\ell\in\mathbb{R}^K$ that have a global order.
Due to this ordering, $\ell$'s associated pairwise differences matrix $\mathbf{C}(\ell):=\ell\mathbf{1}_{1\times K}-(\ell\mathbf{1}_{1\times K})^\top$ is \emph{consistent} \cite{chu1998optimal}, meaning that it is transitive and satisfies cyclic closure $\mathbf{C}_{i,j}+\mathbf{C}_{j,k}+\mathbf{C}_{k,i}=0$.
The unique ranking of consequences may be inappropriate for the data and poorly represent clinical decisions conditioned on present state or behavior.

An alternative approach, inspired by games, infers a more generalized difference matrix directly by comparing \textit{pairs} of observed behaviors, more closely mirroring decision making and its uncertainties in a more general setting. 
An alternative approach, inspired by games, infers a more generalized difference matrix directly by comparing \textit{pairs} of observed behaviors, more closely mirroring decision making and its uncertainties while potentially providing a better fit to the data.
Game theory formally includes concepts such as competition and adaptive choices by rational agents that may not be applicable to J6 without substantial abstraction within a broader system. 
System choices, behaviors, and their consequences are emergent properties, while the observed care process reflects latent physiological, practical, logistical, and legal constraints that are not explicitly represented.
Therefore, observed behaviors may not appear individually rational when governed by collective intelligence \cite{couzin2025collective} arising from provider experience and distributed clinical expertise.
However, the relative consequences underlying these emergent behaviors may still be inferred through a game-based comparison framework \cite{wang2025individual,chandrasekaran2025game}.
Here, static gameplay is solely a mechanism used to contrast J6 behavior pairs (windowed distributions of system states) by their differences in short-time consequences (such as driving pressure, lung compliance, P:F, recruitment energy dissipation, \textit{etc.}).

\subsection{Data sources} \label{sec:sources}
As described in the introduction, each PVS is a coupled biomechanical organism under partial control by the healthcare process (doctors, technicians, insurance companies, \textit{etc}.).
The EHR records choices in patient care including ventilator settings, sedation, posture, medication administration, and patient state. 
Capture of ventilator waveforms and internal diagnostics may exceed the hospital technology infrastructure but can be captured by specialized methods.
Together, this care documentation, ventilator-cycle information, and biomechanical feedback provide information about ventilator management, its circumstances, and the cyborg response. 
However, games can be minimally formulated from interaction and consequence data alone; in this case, the limited collection determines the context.

The clinical data available to this investigation comprise two cohorts.
Both chronicle MV patient encounters within the medical intensive care units (ICUs) at UCHealth University of Colorado Hospital (UCH) recorded with approval by the Colorado Multiple Institutional Review Board (COMIRB).
MV therapy of all recorded patients was provided through the institution's fleet of Hamilton (\url{hamilton-medical.com}) G5 ventilators between 2018 and 2024 during two collection efforts.
Within these data, this work considers only the subset occurring under adaptive pressure ventilation with controlled mandatory ventilation (APVcmv, a proprietary patient-interactive mode) used in UCH ICUs $>$65\% of the time.

\textit{Cohort A} comprises 24 patients who survive their MV encounter following admittance with acute respiratory distress syndrome (ARDS) or risk thereof (including 8 with COVID-19).
They are a subset of 36 encounters recorded under approved COMIRB protocol \#18-1433 for a maximum of 48 hours after intubation.
EHR records are not available so continuous data are limited to: ventilator waveforms of pressure ($p$), volume ($V$ and flow $dV/dt$),   the timeseries of ventilator settings and cycle diagnostics, and static patient information (demographics, anthropometrics, P:F at admission, and outcome). 
Inclusion/exclusion criterion are fully detailed in prior work \cite{sottile2024flow}, which used non-standard esophageal pressure recordings to identify VD. 

\emph{Cohort B} is a more general collection of more than 450 patients collected through an automated pipeline (COMIRB protocol \#20-2160).
Monitor-adjacent technology recorded pressure, volume, and flow signals from critical care infrastructure including ventilators.
These data were subsequently time-aligned with EHR records to compose complete documentation of MV encounters \cite{sottile2025developing}.
Aggregate data comprise static patient characteristics as well as timeseries of: patient posture, medication and therapy administration (including FiO\textsubscript{2}), laboratory measurements (including P\textsubscript{a}O\textsubscript{2} and/or S\textsubscript{p}O\textsubscript{2}, monitored vitals, $p$ and $V$ waveforms, and ventilator settings.
Specific subsets of cohort B extracted by contexts are investigated in applications 2 and 3.

\subsection{J6 representation and comparison} \label{sec:data}  
The clinical data detailed above inform the components of the pairwise-comparison framework, which are sketched here and expanded in referenced passages:
\begin{enumerate}
    \item costs ($q\in\mathbb{R}$) are local measures of MV consequences determined from clinical data linked to VILI or PVS state  (\S\ref{ssec:consequences})
    \item contexts ($c\in\mathcal{C}$) partition the joint PVS-time domain into bins that are, by hypothesis, expected to contain directly comparable states such as patient phenotypes or sub-phenotypes, applied therapies (such as specific ventilator modes), and/or time periods (\S\ref{ssec:contexts}) 
    \item breath categories ($\phi\in\Phi_c$ for $c\in\mathcal{C}$) partition the observable PVS breath space for a given context into a low-dimensional set of classes (\S\ref{ssec:ptypes}) 
    \item strategies ($\pi\in\Pi$) are local breath behaviors quantified by time-windowed statistical summaries of breath types (\S\ref{ssec:strategies}) 
\end{enumerate}
These terms quantify PVS behaviors and MV consequences related through game comparison to empirically identify the payoff matrix $(\mathbf{P})$ (\S\ref{sec:EGT}) within each context separately.

\subsubsection{MV Consequences as Costs}\label{ssec:consequences}
Health-related outcomes and other consequences of MV are extracted from EHR and bedside waveform data sources (electronic health records and pressure/volume waveforms, $p$ and $V$, or collectively $pV$) to define health-impacting costs.
Impacts of MV on patient health at timescales shorter than clinical outcomes are poorly characterized. 
However, certain clinical measurements are linked to VILI through pulmonary mechanics at shorter intervals \cite{bates2018ventilator}.
These observables relate to mechanical forces or their effects such as driving pressure ($p_\text{drive}$, the difference between PEEP and the maximum pressure $p_\text{max}$) and compliance (the ratio of volume to pressure changes, $\Delta V/p_\text{drive}$ or $V_T/p_\text{drive}$  where $V_T$ denotes tidal volume). 
This work uses peak rather, than plateau pressure, because it always exists and is simpler to identify.
In addition to these waveform-derived quantities, changes in P:F and FiO\textsubscript{2}, both measuring alveolar gas exchange efficiency, may indicate VILI.
Measures of MV consequence are termed `costs', where increases indicate worsening state (enforced via a sign change). 

\subsubsection{Contexts}\label{ssec:contexts}  
Contexts partition the cohort PVS data by factors hypothesized to affect care and costs associated with MV.
They aim to stratify the data into comparable sets for analysis and extend stationary patient phenotypes to time-specific patient-care categories. 
Variations arise from patient factors (injury type/severity, history, anthropometrics, stress, co-morbidities), care actions and interventions (ventilator settings, sedation, posture, injury care), and patient-care non-stationarity over time that should be compared separately.
Partitioning data into contexts $c\in\mathcal{C}$ can reduce heterogeneity and stabilize the inference of costs.
Broadly, this means leveraging \textit{a priori} hypotheses to identify subsets comparable through quantified consequences or costs.
Hypotheses can limit time periods being compared as MV costs and care mechanisms are believed to differ at longer timescales \cite{rose2017variation}.
Contexts should also separate ventilator mode, a key categorical variable influencing other factors including the coordinate representation of J6 states.
Ventilator modes may reverse the control-and-response roles of pressure and volume, while hybrid and adaptive modes (like APVcmv) mix them.
States associated with different modes may therefore require different coordinates, making direct comparison impossible.
Beyond separating incomparable data, context definition can refine sub-cohort definitions, treatments, or time periods to identify HTEs (\S\ref{HTEs}).
An example of a context is: data from the first 3 days of MV of male patients occurring at low PEEP settings ($\le 5$ cm\,H\textsubscript{2}O) in volume-control ventilation modes without neuromuscular blockade (NMB, a paralytic).

\subsubsection{Breath Categorization}\label{ssec:ptypes}  
Ventilation waveform data constitute high-dimensional continuous features that can be approximately discretized into breath-event sequences.
Longer time J6 analysis benefits from further dimensional reduction.
Breath types can be defined by empirical feature similarity or by VD typology \cite{sottile2018association,sottile2020ventilator}.
Here, MV breath types define categories $\Phi = \{\phi_k\,|\,k=1..K\}$, reducing continuous-time PVS data to breath-type sequences.
For the $n$th MV data sample, define the state sequence $S_n = \{s_n[j] \mid j \in T_n\}$, where $s_n[j]$ denotes system states and $T_n = \{1..J_n\}$ indexes the $J_n$-length timeseries.
The raw data $s_n$ consist of $pV$ waveform data discretized into sequences of breath behaviors plus ventilator settings defining PVS states.
Each state $s_n[j]$ is labeled by one of $K$ low-dimensional classes $\Phi$, referred to as breath types, based on the structural similarity of waveform and ventilator-settings features.
Waveform parametrization via dummy-model inversion and similarity-based categorization of MV data are detailed in \cite{stroh2023hypothesis} and \cite{stroh2024identifying}, respectively. 
Trajectories $S_n$ are expressed by the indices $\{f_n[j]\}_{j=1}^{J_n}$ of state-approximating categories: $\{s_n[j]\} \approx \{\phi_{f_n[j]}\}\to \{f_n[j]\}$.
Category sets may be context-indexed as $\Phi_c=\{\phi_k^c\}_{k=1}^{K_c}$, where $K_c$ indexes the context-specific breath types.

\subsubsection{Strategies}\label{ssec:strategies} 
MV consequences evolve over multi-hour timescales spanning hundreds of breaths. 
This timescale separation justifies treating temporal order within each analysis window as negligible, allowing windowed breath-category sequences to be represented by frequency statistics. 
These summaries are called \textit{strategies} for their role in games, which compare strategies based on the differences in their associated consequences. 
For a sequence of breath types $S_n$ over some window, the vector of type occurrence percentages $\pi_n$ lies in the probability simplex $\Delta^{K-1}=\{ \sum_{k=1}^K \alpha_k \phi_k \mid 0\le\alpha_k\le1, \left\lVert\alpha\right\rVert_1=1 \}$ (where $K=|\Phi|$ is the number of MV breath types).
The data yield observed strategies $\Pi := \{\pi_n\}$, a sample from $\Delta^{K-1}$.

\section{Game-inspired Consequence Attribution via Pairwise Comparison}   \label{sec:EGT} 
Games are adopted here to model pairwise interaction among observed strategy--cost pairs $\{{\pi_j, q_j\}}$ under an unknown payoff. 
Unlike formal game theory, it does not model strategic choice and behaviors, adaptation, or equilibrium under a known payoff process.	
Two-player games compare strategies $\pi\in\Delta^{K-1}$ using a payoff function $\mu: \Delta^{K-1}\times\Delta^{K-1}\to\mathbb{R}$ which determines the results of the game for each player.
In normal-form games, $\mu$ determines payoffs for the interaction of strategies $\pi_i,\pi_j\in\Pi$ via a payoff bi-matrix $(\mathbf{B}_1,\mathbf{B}_2)$ whose elements define payouts for players 1 and 2, respectively.
Total outcome payoff is then determined by the difference in individual payouts: $\mu(\pi_i, \pi_j) := \pi_i^\top \mathbf{B}_1\pi_j - \pi_i^\top \mathbf{B}_2 \pi_j$. 

Formulated under the assumptions of zero-sum gameplay and player equivalence ($\mathbf{B}_1 = \mathbf{B}_2^\top$), the resulting payoff matrix $\mathbf{P}:=\mathbf{B}_1-\mathbf{B}_1^\top$  is skew-symmetric and maps strategy pairs to payoff differences: 
\begin{equation} 
	q_i - q_j = \mu(\pi_i,\pi_j) =  \pi_i^\top\mathbf{P}\pi_j.
\end{equation} 
Here, coordinates of $\mathbf{P}$ correspond to MV breath types while $\mu$ gives the difference in \emph{cost} between the strategies.
Consequently, $\mathbf{P}_{i,j}>0$ indicates that breath type $\phi_i$ is costlier than $\phi_j$ and therefore a less preferable option.

\subsection{Inferring the Payoff from Comparisons of J6 Data} 
Within each context, games link empirical J6 behaviors to their consequences through an unknown payoff process that can be inferred from interaction among strategy--cost data pairs. 
The payoff process is assumed to be linear so its matrix $\mathbf{P}\in \mathbb{R}^{K\times K}$ contains the pairwise consequence differences of the $K$ breath categories (\textit{i.e.}, the pure strategies). 
Identifying the payoff matrix is the following inverse problem:
Given examples $\{(\pi_i,\pi_j), q_i-q_j\}$ of the game, find the unknown payoff table $\mathbf{P}$ so that $\mathbb{E}[\mu(\pi_i,\pi_j)]$ approximates cost data $q_i-q_j$ across all $i,j=1..N$. 
The solution will depend on various factors, including:
\begin{enumerate} 
  \item how breath categorization and windowing define state and strategy spaces
  \item which costs are computed from clinically-relevant observable outcomes
  \item how contexts partition the cohort population of strategies
\end{enumerate}

In the practical problem, cost $q$ associates with a strategy $\pi$ derived from state category sequence data: $S_i\mapsto q_i$ for some cost function. 
The game defines a method for relating sequences $S_i,S_j$ to differences in observable outcome involving the unknown payoff matrix:
\begin{equation}
        \begin{tikzcd}[column sep=0.45cm]
        S_i\arrow[r,"\text{d}"] \arrow[rdd, swap,"\text{cost}"] & 
          \pi_i\arrow[rd, swap,dashed] 
          & & \arrow[ld,dashed ]\pi_j
          &
          \arrow[l,swap,"\text{d}"] S_j \arrow[ldd, "\text{cost}"] \\
          & & \fbox{ $q_i-q_j = \pi_i^\top\mathbf{P} \pi_j$} & &\\
           & q_i \arrow[ru,dashed ]\ & &  \arrow[lu,dashed ]q_j 
        \end{tikzcd}
          \label{eq:game}
\end{equation}
Here, $S_i,S_j$ are samples of observed breath type sequences, $\pi_i,\pi_j$ are their respective associated strategies, $q_i,q_j$ are externally measured/observable costs/rewards of those behaviors, and $\mathbf{P}$ is the payoff matrix. 

\subsection{Estimating Payoff Matrices via Regression} \label{sec:EGTinv} 
Identifying relative payoffs is equivalent to solving the inverse problem $q_i-q_j = \pi_i^\top\mathbf{P} \pi_j$ from the population of $N$ examples $\{(q_n, \pi_n)\} \subset \mathbb{R} \times \Delta^{K-1}$. 
Symbols $x,y$ replace indexed $\pi$'s below to distinguish left and right strategies with respect to cost differences.
That is, a new population of examples $\{(\Delta q_m,x_m,y_m)\}_{m=1}^{M}\subset\mathbb{A}^1(\mathbb{R})\times \Delta^{K-1}\times \Delta^{K-1}$ simply renames $\{(q_i-q_j,\pi_i,\pi_j)\}_{i,j=1}^N$ for notational convenience. 

The cost difference of the $m$th comparison is a Frobenius inner product:
\begin{align}
  \Delta q_m = x_m^\top\mathbf{P} y_m = \langle x_my_m^\top, \mathbf{P} \rangle_\text{Frob} =w_m^\top \mathbf{p}
\end{align}
via Kronecker product identities \cite{van2000ubiquitous}.
Here, $w_m:= \text{vec}(x_my_m^\top)\in\mathbb{R}^{K^2}$ and $\mathbf{p}:=\text{vec}(\mathbf{P})\in\mathbb{R}^{K^2}$ are the vectorized strategy outer product and unknown payoff matrix, respectively.

However, the skew symmetry reduces the number of free parameters in $\mathbf{p}$ while respecting simplex geometry requires that $\mathbf{P}\mathbf{1}_K$ vanish.
The matrix $\mathbf{D}\in\mathbb{R}^{K^2\times K_p}$ mapping $\mathbf{p}$ to constrained free parameters $\mathbf{\beta}\in\mathbb{R}^{K_p}$ is trivial to construct and  re-parametrizes the problem as
$\Delta q_m = w_m^\top\mathbf{D\beta}$.
The $M$ scalar equations $\Delta q_m = w_m^\top \mathbf{D}\mathbf{\beta}$ 
yield the over-determined system $\mathbf{\Delta q} = \mathbf{W^\top D}\mathbf{\beta}$ where $\mathbf{\Delta q} \in \mathbb{R}^M$ contains the cost differences $\Delta q_m$ and $\mathbf{W} \in \mathbb{R}^{K^2 \times M}$ has columns $w_m$.

The solution $\widehat{\mathbf{\beta}}$ identifying $\mathbb{E}[\mathbf{\Delta q}\,|\,\mathbf{W}]$ is found by regression or optimization, and then reshaping the vector $\mathbf{D}\widehat{\mathbf{\beta}}$ into a matrix to estimate $\mathbf{P}$. 
This work uses $\mathbf{P}$ throughout to denote both the true and estimated matrix except in verification tests where the true payoff is known.

The entries of $\mathbf{P}$ quantify relative consequences between breath types rather than a global cost ordering. 
Consequently, breath-type costs can be recovered only up to a constant reference value.
Also, $\text{rank}(\mathbf{P})$ may exceed 2, whereas consistent difference matrices are always rank 2. 
Therefore, the inferred payoff need not admit unique (up to translation) or globally orderable breath-type costs.
		
\section{Numerical Experiments} \label{sec:results}
This section begins with synthetic verification experiments using games to identify payoff tables over breath types.
Compared with the baseline result (Experiment 1), recovering the payoff structure from strategy--cost pairs fails for breath types linked to multiple distinct costs (Experiment 2). 
In contrast, portions of the payoff matrix are accurately estimated where breath types have unique and consistent costs(Experiments 2 and 3).
These results help establish confidence in clinical applications, where breath type-cost links are unknown.	
Later, applications 1--3 apply game-based inference to clinical data to find MV costs of ARDS patients, differences in MV cost based on ARDS status, and MV cost changes over time.

\subsection{Verification}  \label{sec:verify}
The game framework relates strategy pairs to a single cost difference (Eqn. \ref{eq:game}) so it cannot meaningfully generate simulated strategy--cost pair data from a given payoff matrix.
Instead, test data were generated by assigning a vector of costs $\ell_1$ over breath types.
This vector defines a consistent payoff matrix $\mathbf{P}=\mathbf{C}(\ell_1)$ as well as true sample costs $q_n=\ell_1^\top\pi_n$ for any strategy $\pi_n\in \Delta^{K-1}$.
A population of data $\{(\pi_n,q_n)\}$ are then generated by non-uniform random sampling and paired with noisy cost values. 
Experiments use these synthetic strategy--costs pairs generate data and evaluate the effects of breath categorization uncertainties.
Because both breath types and associated costs are clinical unknowns, it is important to consider what can be learned about costs when assumed breath categorization is incorrect.
Experiment 1 is a baseline test of payoff estimation informed by noisy synthetic data in the case where breath types coincide with the cost-equivalence classes of breaths. 
Experiments 2 and 3 modify Experiment 1 to examine the effects of payoffs inferred on under- and over-specified breath types, respectively.

\subsubsection{Experiment 1}  
\begin{figure}[!htb] 
    \centering
    \includegraphics[width=.7\textwidth]{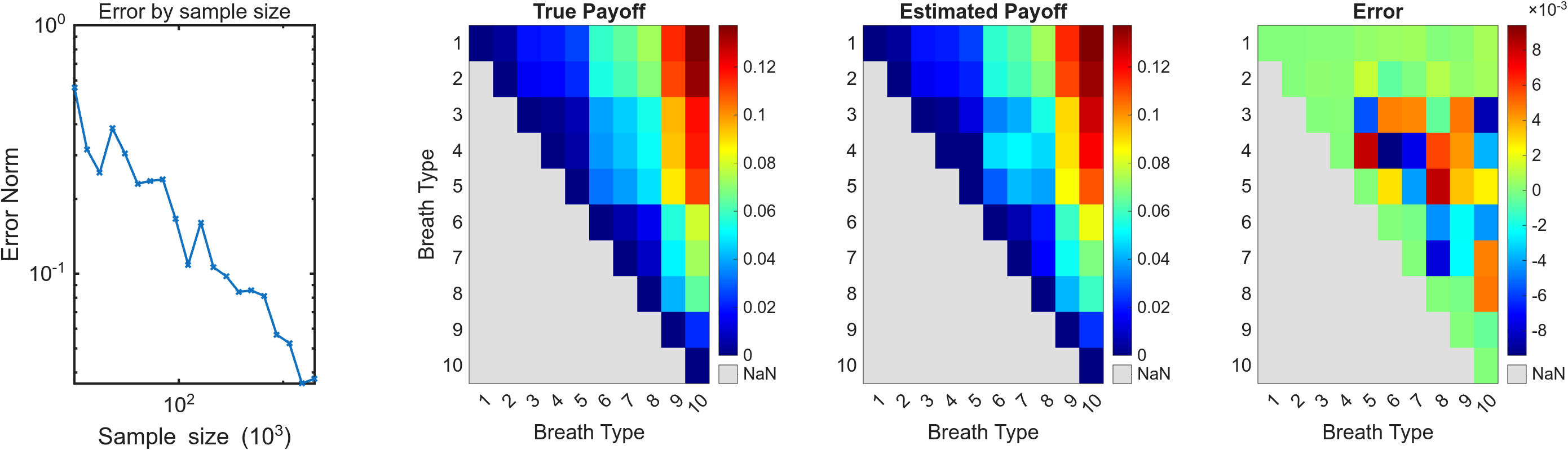}  
    \caption{The baseline estimation results, Experiment 1.    
    Strategies are sampled from $K=10$ breath types sorted in increasing cost and decreasing frequency, while cost data for breath types are polluted with 3\% additive random noise when sampled.
    Rare-vs-rare type comparisons (those among types with higher index) require a higher number of samples to accurately resolve breath type-cost associations.    
    The left plot shows decrease with respect to sample size for Frobenius norm errors between the estimated payoff matrix $\mathbf{\widehat{P}}$ and true payoff matrix.    
    The values in the upper triangle of the true payoff matrix (`True'), the lowest-error estimated payoff matrix (`Estimated'), and the difference between them (`Error') appear as the heatmaps in the remaining plots.
    Payoffs heatmap values correspond to row-minus-column breath type cost differences.
    }
    \label{res:ex1}
\end{figure}

In this baseline experiment, the global distribution on $K=10$ breath type occurrences is assumed to decrease exponentially from 35\% for $\phi_1$ to 2\% for $\phi_{10}$, while associated costs-per-breath rise exponentially from 0.85 to $\sim$5. 
These breath types and their costs are also assumed to be the ground truth in experiments 2 and 3.
Figure \ref{res:ex1} panels show the decay of error rate with increasing sample size as well as heatmaps of the upper triangle of payoff values.
Three-percent random noise is added to sampled costs (prior to subtraction); RMS errors in the estimated payoff matrix are uniformly less than 5\% of the true value (the scale of noise added to costs) even for rare events after using about 1 million sample game instances (\textit{i.e.}, $10^3$ kilosamples of strategy interaction).
The result suggests very little solution bias provided the population of strategy--cost data is sufficiently large.

\subsubsection{Experiment 2}
\begin{figure}[!htb] 
    \centering
    \includegraphics[width=.7\textwidth]{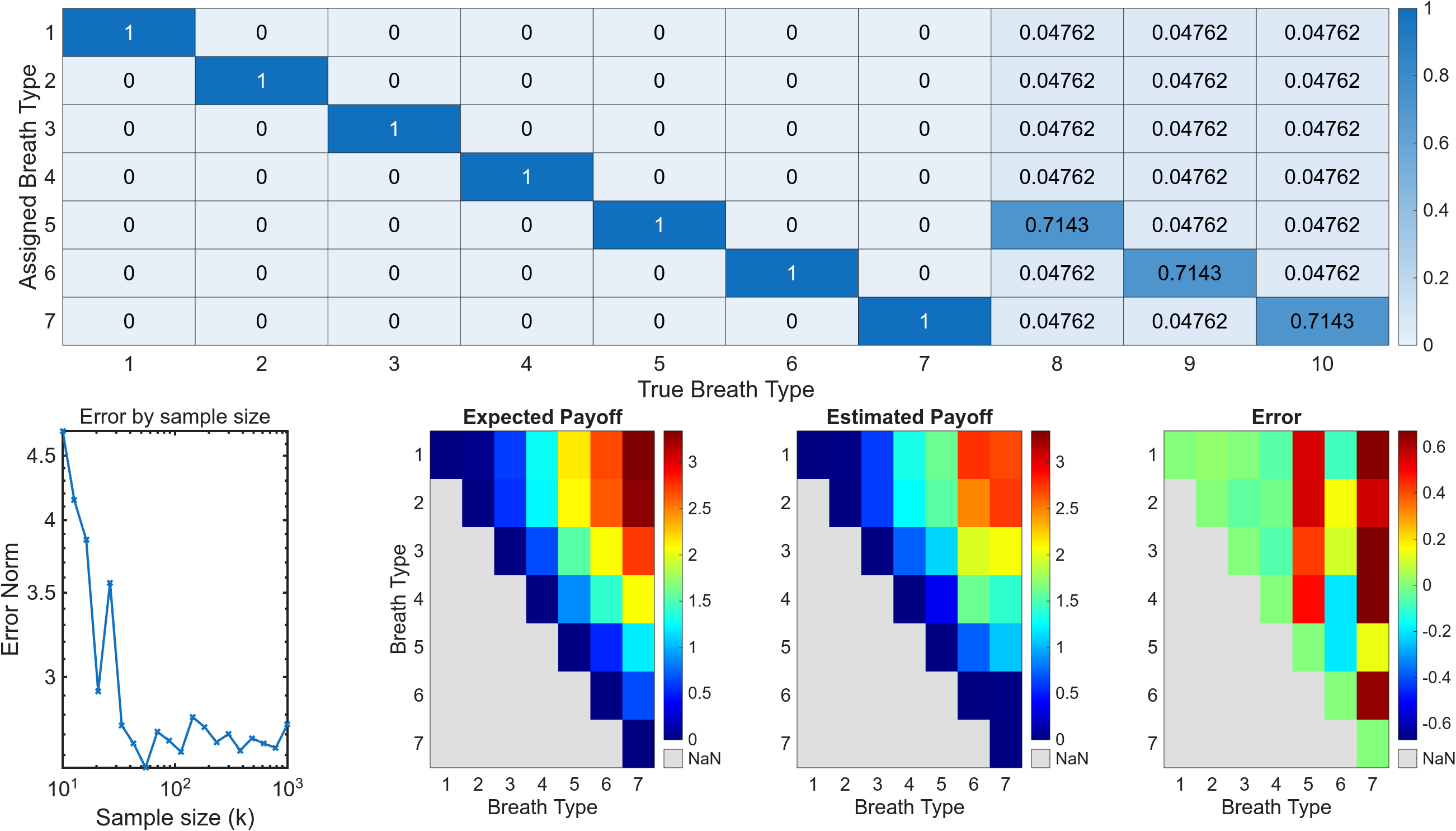} 
    \caption{Effects of under-specified breath types, Experiment 2.(\textit{top}): 
    The upper row shows the heatmap of the matrix used distribute $K=10$ cost-distinct breath types into 7 assumed breath types. 
    True breath types are sorted in increasing cost and decreasing frequency, but only the first 7 are correctly specified. 
    The last 3 types (8,9,10) are the rarest also most costly/injurious/detrimental breath types but are misidentified disproportionately across the first 7 types.
    (\textit{bottom}): 
    Panels parallel the lower panels of the previous figure, showing, left-to-right, Frobenius norm errors \textit{vs.} sample size, and heatmaps of expected payoff matrix given reassignment probabilities, the  approximation based on the largest sample size, and the errors.
    Errors reduce but quickly stagnate; the expected payoff matrix cannot be accurately estimated by increasing the sampling.
    }\label{res:ex2}
  \end{figure}

This experiment examines the effects of an incorrect breath categorization in which the 10 cost-differentiated types of the previous example are hypothesized as 7 types.
It represents a noisy partition of the breath space coarser than cost equivalence classes.
To create this misalignment, breaths of three breath types are randomly re-assigned in strategy windows to other in a non-uniform way: types 8--10 are randomly redistributed over types 5--7, respectively, with probability 15/21 and to one of the other 6 types with uniform probability 1/21.
Results (Fig.\ref{res:ex2}) show that mis-specified or insufficiently granular breath types result in large errors that increased sampling size cannot overcome. 
However, when similar relabeling only corrupts types 5--7, the cost differences among types 1--4 are identified correctly.

\subsubsection{Experiment 3} 
\begin{figure}[!htb] 
    \centering
    \includegraphics[width=.7\textwidth]{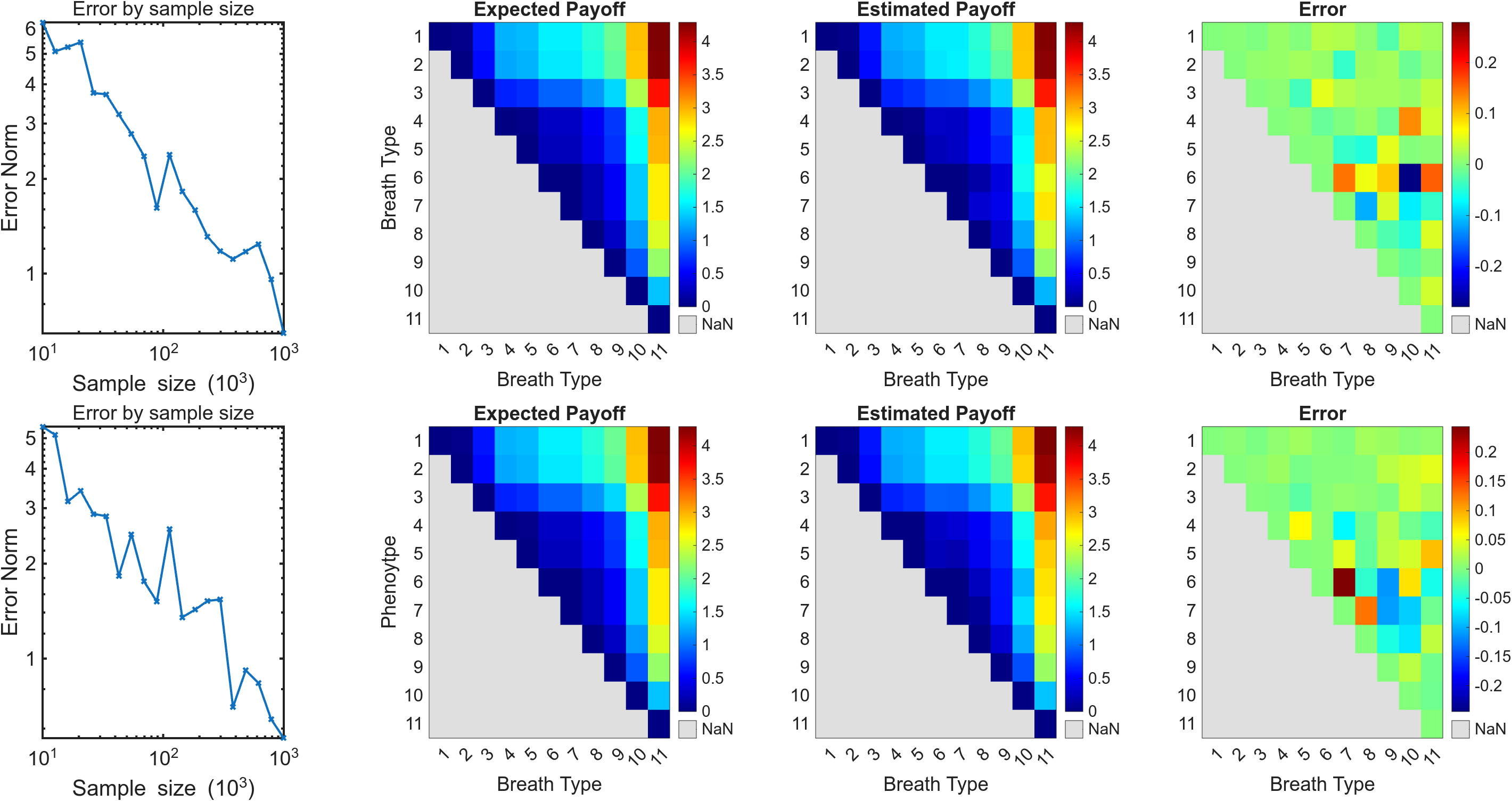} 
    \caption{Effects of over-specified breath types, Experiment 3.
    The layout for each row of panels follows that of the bottom row of Fig.\ref{res:ex2}.
    (\textit{top}): Contents of type \#6 of intermediate cost and frequency are randomly assigned to create cost-identical types \#6 and \#7. 
    (\textit{bottom}): Type \#6 is randomly reassigned entirely to type \#7 with probability 0.7.
    Both assignments yield underestimates of the costs of these breath types while finding their costs to be the same ($\mathbf{P}_{6,7}\approx 0$).
    The errors are generally ---but not exclusively--- localized to payoff entries associated with these types, and decrease rather than stagnate with increasing sample size.
    }\label{res:ex3}
\end{figure}

MV breath space partitioning may also define more breath types than cost-equivalence classes by over-specifying breath categories.
The situation is modeled by duplicating one of the 10 cost-distinguished types of Experiment 1 and inferring the payoff over 11 breath types.
Specifically, a new type \#7 duplicates type \#6 and is populated by randomly assigning breaths from type \#6 to it (with original types \#7--10 relabeled as \#8--11).
Figure \ref{res:ex3} shows two reassignment processes: type \#6 strategy components are randomly split between equivalent labels \#6 and \#7, or randomly labeled entirely as one or the other.
In both cases, errors are generally localized in the duplicated types with the payoff correctly recognizing them to be nearly cost-identical ($\mathbf{P}_{6,7}\approx 0$).
Errors in $\mathbf{P}$ involving the other types persist even for large sample sizes for all but the most commonly occurring types.
Full mis-labeling produces larger errors than the partial mis-labeling,  particularly for payoff entries involving rarer types due to reduced sampling.

\subsection{Applications to clinical data}
Formulating games on clinical J6 data requires application-specific definitions, including the comparison context and breath categories representing those data. 
The comparison context is selected according to the purpose of each application, while corresponding J6 data are categorized (\S\ref{ssec:ptypes}) following prior work \cite{stroh2024identifying}.
Briefly, breath categories are obtained by clustering low-dimensional representations (\cite{mcinnes2018umap}) of parameterized pressure and volume waveforms \cite{stroh2023hypothesis} together with selected ventilator settings (PEEP, tidal volume, respiratory rate, and inspiration-to-expiration ratio).
Details of the data, context, categories, and windowing defining strategy--cost pairs are provided within each application below.

The length of matrix $\mathbf{W}^\top$ is quadratic in the number of strategies and can exceed available memory for clinically relevant datasets. 
Consequently, payoffs from each context's data are calculated by Monte Carlo estimation over large random subsets of strategy pairs rather than the full population.
Stable estimation requires each subset to sufficiently cross-sample breath categories so $\mathbf{W}$ constrains every row (and column) of $\mathbf{P}$. 
Rare and poorly resolved categories are therefore combined into a single heterogeneous category, while repeated random sampling provides adequate coverage of the remaining categories.

\subsubsection{Application 1: Exploring MV costs of ARDS patients} \label{rwd_app1}
The game-based analysis applied to real-world clinical data of cohort A (\S\ref{sec:sources}) explores the MV costs of ARDS patients for whom respiratory management is an urgent and important concern.
Data from these 36 ARDS or ARDS-risk patients with esophageal balloon further omit: the first 6 hours on records when patients are typically paralyzed following intubation and/or esophageal balloon placement, patients who died during the MV encounter, and periods \emph{not} using the APVcmv ventilator mode. 
The context ($c$) frames the $\sim$620 hours of recorded data from 24 MV patients meeting these criteria.

These data were segmented into 222K 10-second disjoint intervals and grouped by feature similarity to identify 29 breath types ($\Phi$).
Eight types, each associated with less than 1\% of data, were omitted because they suggest anomalous data or extremely rare behaviors.
The remaining $K=21$ types captured more than 95\% of the data framed by the context.
Game-inspired inference (\S\ref{sec:EGTinv}) identified payoff matrices linking 5-minute strategies (\textit{i.e.}, occurrence distributions) to costs defined by 1-hour mean driving pressure ($p_\text{drive} := \max(p) - \text{PEEP}$) measured 6 hours later.

\begin{figure}
  \centering
 \includegraphics[width=.9\textwidth]{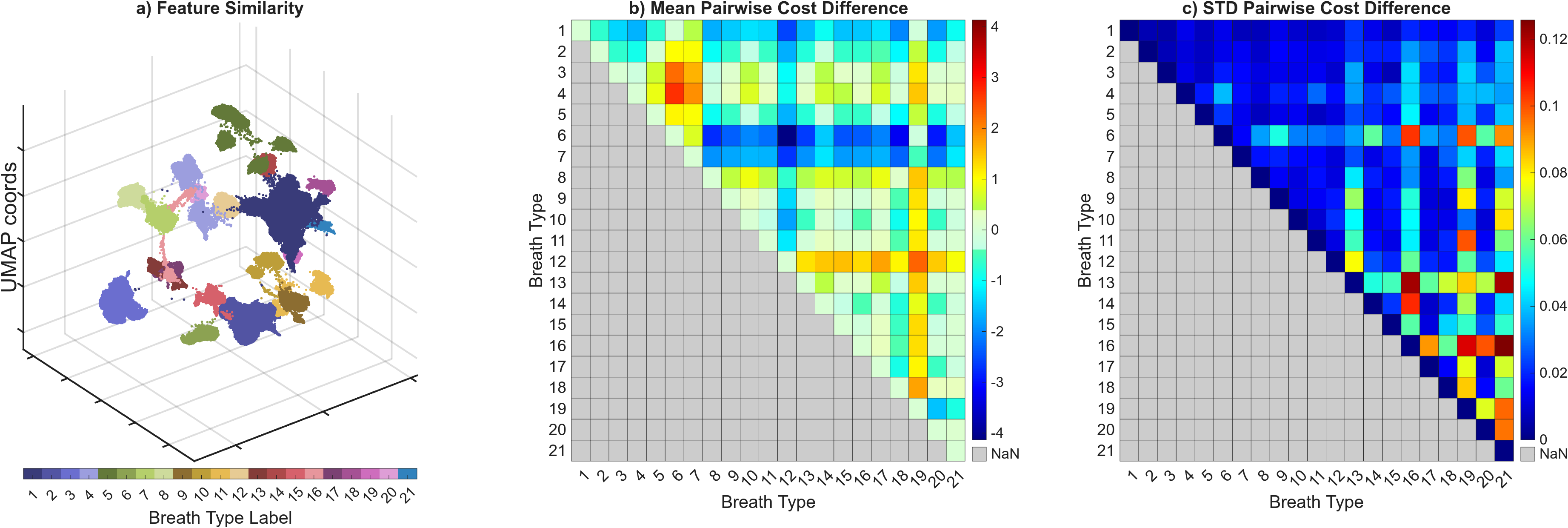}
  \caption{\textit{(left)} Similarity-based breath types of contextualized breath data, \textit{(center)} mean estimates of comparative costs based on driving pressure ($p_d := \max(p)-\text{PEEP}$), and \textit{(right)} standard deviation estimates of comparative costs for clinical human data. 
  At left, UMAP reveals the global and local similarity structure used to categorize J6 data, with each point representing a 10-second interval of patient data.   
  Only relative point locations are shown because UMAP interpretation depends on relative proximity rather than the coordinate values.  
  In center and right, heatmaps of $\mathbf{P}_{i,j}$ show driving pressure cost differences between breath types $i$ and $j$.
  Breath types are ordered by frequency within the population, with payoff entries involving rarer types generally exhibiting greater uncertainty.
  }
  \label{fig:mainREDO}
\end{figure}

Figure \ref{fig:mainREDO} shows type-labeled data in Uniform Manifold And Projection (UMAP, \cite{mcinnes2018umap})  coordinates (left, labels ordered by volume) alongside heatmaps of payoff matrix statistics. 
This categorization was used to represent patient-level J6 timeseries, from which windowed occurrence summaries and lagged pressure data were used to construct strategy--cost pairs over the cohort.
Following (\S\ref{sec:EGTinv}), the payoff matrix mean (center) and standard deviation (right) were computed from 10 bootstrap approximations,
each estimated using independently drawn 80\% subsamples of the observed strategy--cost pair population for the left and right comparison sets.

\paragraph{Payoff Interpretation and Cost Ordering of Breath types}
Recall that $\mathbf{P}_{i,j}>0$ indicates that type $i$ (row index) is costlier than type $j$ (column index).
The values of $\mathbf{P}$ represent edge weights in an anti-symmetric directed graph with edges pointing toward the costlier node.
This suggests the use of a graph to identify and compare cost orderings through Hamiltonian paths using edges all of one sign.
Finding the minimum-cost Hamilton path over $\mathbf{P}$ is NP-hard (e.g.,\cite{garey1979computers,korte2011combinatorial}), but a reasonable ordering is the nearest-neighbor path from best to worst types (identified by highest out- and in-degree, respectively) over positively weighted edges.
Another global cost is associated with the inferred payoff's nearest consistent approximation \cite{chu1998optimal,bozoki2008solution},  $\ell_*:=\text{arg\,min}_\ell \left\lVert\mathbf{P} - \mathbf{C}(\ell)\right\rVert_\text{Frob}^2$.

\begin{figure}[!htb]
    \centering
    \includegraphics[width=.75\textwidth]{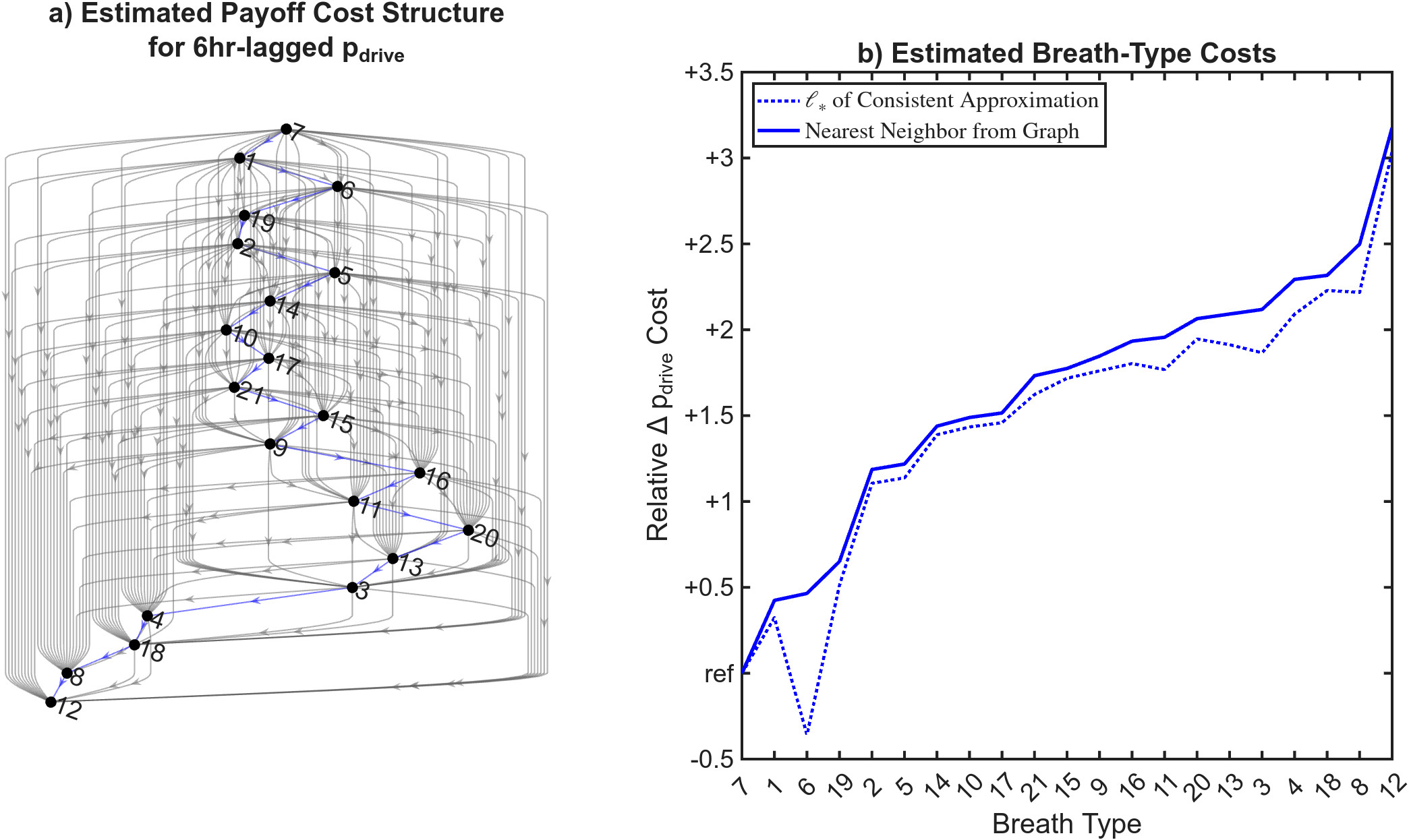}      
    \caption{Breath type cost-rank by adjacency graph (\textit{left}) and its comparison with the consistent cost approximation (\textit{right}).
    Note that cost estimates are means and triangle equality is not generally satisfied by the mean.     
    For example, in the mean payoff, $\mathbf{P}_{18,8}+\mathbf{P}_{8,12}=0.8535<1.0625=\mathbf{P}_{18,12}$ with similar features in each Monte Carlo iterate.
    These triangle equality violations are not explained by estimated uncertainties in payoff values and indicate $\mathbf{P}$ is inconsistent with a global breath type-cost vector.
    The nearest neighbor solution aligns with the nearest consistent solution except at type \#6.
    }
    \label{fig:orderedcosts}
\end{figure}

Figure \ref{fig:orderedcosts} compares the breath type costs determined by the nearest neighbor path across the directed graph and by consistent approximation. 
Ranking in the later differs in that breath type \#6 has lower cost than \#7 in the consistent approximation.
Pairwise distances within the payoff prevent a unique linear cost ordering, which would require transitivity (\textit{i.e.}, cyclic closure $\mathbf{P}_{i,j}+\mathbf{P}_{j,k}+\mathbf{P}_{k,i}=0$).
Breath type costs are not uniformly distributed and those labeled \#7 and \#12 are clearly toward the best and worse ends.
The distribution is left-skewed and includes a cluster of types with moderately low cost (skewness $-0.41$) while larger differences among the costlier types correspond to normal tail behavior (kurtosis $3.04$).
Reconstructed cost orderings are based on point estimates of $\mathbf{P}$, although payoff uncertainty could be incorporated in future work.

\begin{figure}[!htb]
    \centering
    \includegraphics[width=.7\textwidth]{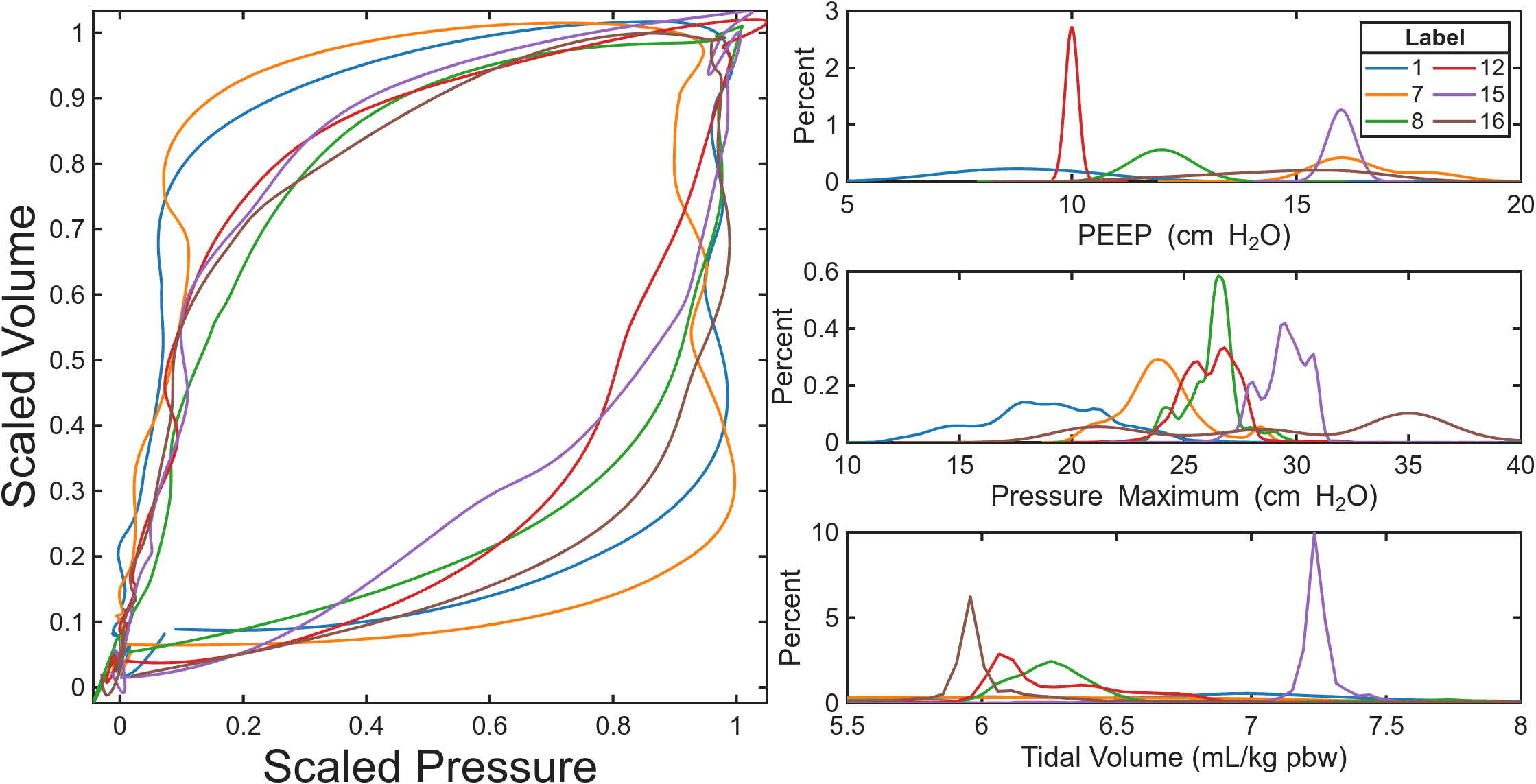}
    \caption{Waveform (as normalized $pV$ loops) and vent settings characterizations associated with breath types labeled 7 (best), 1, 15, 16, 8, and 12 (worst). 
    Fig.\ref{fig:orderedcosts} orders them by increasing cost as $\phi_7<\phi_{1}<\phi_{15}<\phi_{16}<\phi_{8}<\phi_{12}$.
    Insights into clinical aspects of MV hinge on interpreting these breath properties in relation to consequences such as those inferred by game-inversion.
    } 
    \label{fig:char}
\end{figure}

\paragraph{Characteristic Interpretation}
Figure \ref{fig:char} characterizes data from several breath types including normalized pressure-volume ($pV$) loops near the group median in UMAP coordinates and distributions of key $pV$ properties.
With respect to the cost of driving pressure measured 6 hours later, the least cost types are characterized by different PEEP settings that yield relatively low maximum pressure through appropriately set tidal volume. 
Their \textit{current} driving pressures (\textit{e.g.}, 7.42 and 10 cm\,H\textsubscript{2}O for \#7 and \#1, respectively) contrast the most costly types, which associate with high maximum pressures (\textit{e.g.}, 14.25 and 16 cm\,H\textsubscript{2}O for \#8 and \#12, respectively).
The $pV$ loop of costliest type, \#12, also features the occurrence of `beaking' in late inspiration, suggesting over-distention resulting from inappropriately high maximum driving pressure. 
Although driving pressure does not define types, cost rank based on its 6-hour future value aligns qualitatively with the current-time driving pressure order derived from type characteristics.

\subsubsection{Application 2: Exploring patient heterogeneity} \label{rwd_app2}
Application of game-based analysis to the data of cohort B examines breath-cost differences within a larger,  more diverse patient set during the initial period of MV.
The data considered are the 6--30 hour period of 3--7 day MV encounters, restricted to use of APVcmv mode while the patient was in semi-Fowler's position (the most common in these data).
The context isolates a data subset of approximately 4.1 million breaths for 329 patients, and these data were coarsely categorized into 26 basic types. 
For simplicity, the most common 10 types (each $>3.5\%$ by volume) are used plus an 11th category designated as `999' aggregating the remainder (each $<=3.1\%$, totaling $\sim$30\% of the data).
The costs of MV for these breath types are explored in two contexts differentiated by the presence or risk of ARDS following \S\ref{sec:EGTinv} as in previous applications.
Type occurrence distributions (strategies) over disjoint $\sim$30-minute intervals are hypothesized to associate with 4-hour average P:F ratios recorded 24 hours later (costs).  

\begin{figure}[!htb]
    \centering
    \includegraphics[width=\textwidth]{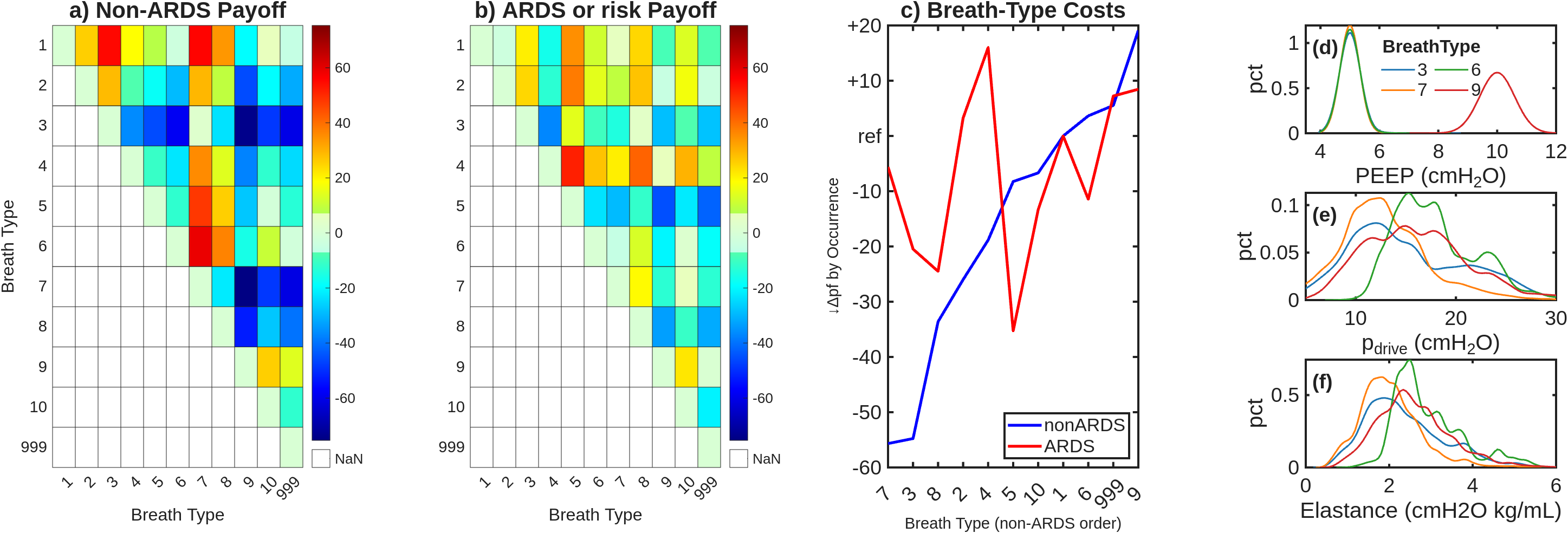}
    \caption{Payoff matrices for breath types during hours 6--30 in patient cohort B. 
    Payoff matrices are for patient phenotype differentiated contexts: non-ARDS patients \textit{(a)} and those with ARDS or identified risk \textit{(b)}. 
    Differences correspond to 24-hour decreases in 4-hour mean P:F ratio so that positive cost indicates that the row type is associated with worse gas exchange than the column type (\S\ref{ssec:consequences}).
 	 Breath type cost ranking (\textit{c}, via nearest consistent cost) depends on the interaction between patient properties, ventilator settings, and their interaction (\textit{d--f}).
	 Costs in panel \textit{c} are aligned at breath type \#1, although its absolute cost likely differs between the two groups.
    The results show a clear difference in both the payoff structures and cost ordering, indicating that the relative consequences of breath types depend on patient condition.
    The three right-hand panels (\textit{d--f}) summarize aspects of data associated with breath types to aid interpretation of these differences and are not derived from the inferred costs.    
    }
    \label{fig:rwd2}
\end{figure}

Figure  \ref{fig:rwd2} compares the payoffs (\textit{a,b}) and associated nearest-consistent cost orderings inferred from P:F ratio decreases for the ARDS-stratified contexts (\textit{c}) with additional panels (\textit{d--f}) showing breath type characteristics to aid interpretation.
The non-ARDS subcohort (107 patients) began MV without ARDS or associated risk factors, whereas the ARDS subcohort (222 patients) did. 
Payoffs were estimated from 1000 bootstrap replicates, each drawing $\sim$256K strategy--cost pairs with replacement from the population of $\sim$14M possible pairs. 
Cost distributions clearly differ between the groups, with larger decreases in P:F (higher costs) generally associated with breaths exhibiting higher driving pressures and elastance. 
Likewise, the breath-type cost orderings also differ between the ARDS and non-ARDS contexts, indicating that the relative consequence of a given breath type depends on the patient's clinical condition rather than a universal ordering. 
Costs differ between these results and sex-stratified inference (SI Fig.\ref{fig:null_and_sex_app2}) for the same data, indicating context-dependent cost order and structure.

The non-ARDS context features a wider range of costs than the ARDS one, possibly reflecting the more intensive clinical management provided to the more vulnerable ARDS group. 
Assuming a common cost for breath type \#1 in both contexts (\textit{c}), the bundle of rare types (\#999) also appears to have equal costs, although the frequencies also differ in cohort strata.
In terms of notable characteristics (SI \ref{SI:rwd_app2_ptypes}), breath types least effective at increasing P:F ratios,  \#6 and \#9, associate with relatively high driving pressures and high elastance.
This is qualitatively consistent with the rationale underlying the ARDSNet protocol.
Reduced P:F gains from higher driving pressures in stiff lungs help justify the ARDSNet lung-protective strategy of lower tidal volumes. 
However, the same high PEEP/low tidal volume strategy may be less favorable in non-ARDS patients with low lung elastance. 
These results suggest patient respiratory health factors such as elastance influence observed cost differences crucial to optimizing oxygenation-cost tradeoff.

\subsubsection{Application 3: Exploring MV consequence evolution} \label{rwd_app3}
The third application examines longitudinal in changes breath-type consequences by inferring payoffs from moving windows of data for a fixed set of patients.
The application considers the initial 0--72 hour period of MV 3--7 day encounters in cohort B to ensure patient persistence throughout the experiment while excluding prolonged MV encounters whose early care may differ.
From the first three days of cohort B, only breaths under APVcmv ventilation while patients were in the semi-Fowler's position were retained and 20 breath types were identified for these data (SI \ref{SI:rwd_app3_ptypes}).

\begin{figure}
	\centering	
	\includegraphics[width=.75\textwidth]{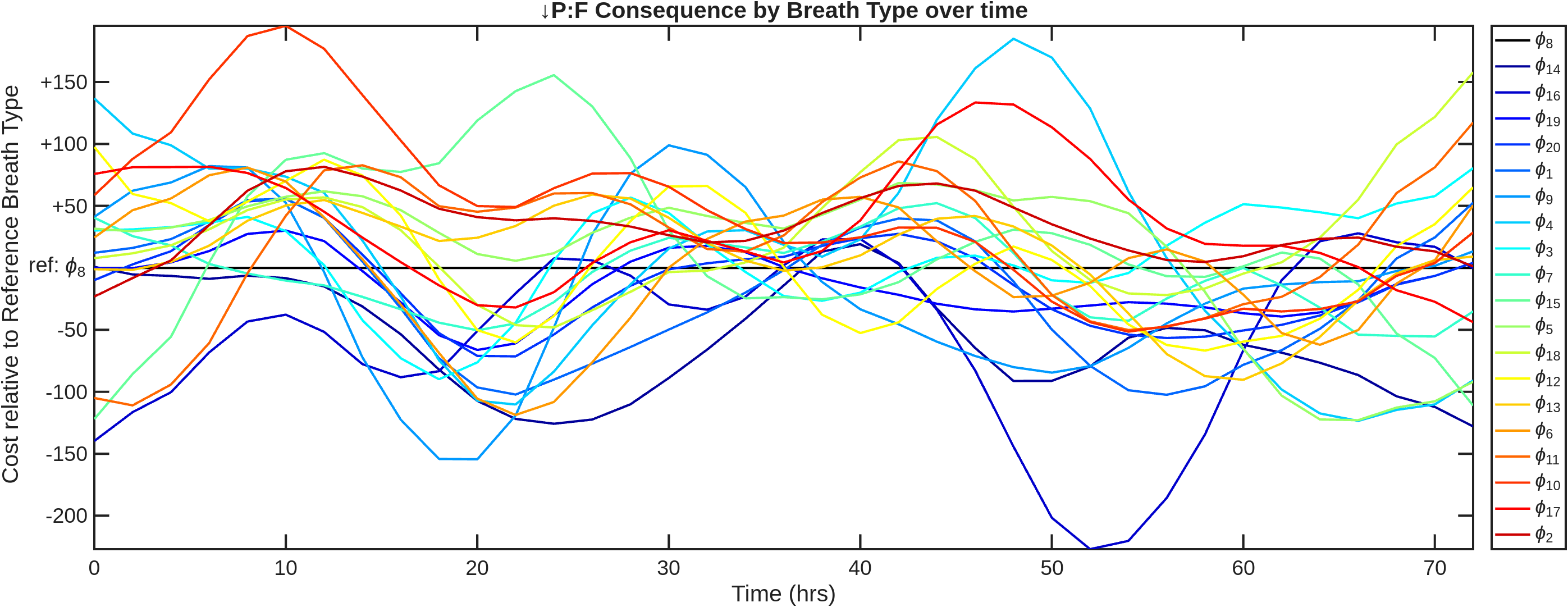}
	\caption{Time evolution of breath costs within a moving context. 
	Panels show payoff-derived breath type costs computed in 6-hour context windows associating decreased P:F ratio with $\sim$1200 breath strategies.
	Costs at each time are depicted relative to type \#8, which was identified as nearest to the mean temporal cost over all types and a suitable choice as a persistent reference. 
	The legend orders types from least to most injurious, on average in relation to type \#8.
	For uniformity, types underrepresented in some window (\textit{i.e.}, those bundled into \#999) are hidden.
	The global structure of costs may be driven by common care practices, such as downward titration of FiO\textsubscript{2} after 12 hours of MV.
	}
	\label{fig:costseries}
\end{figure}

Applying the game-inspired analysis on 6-hour disjoint moving windows, payoff matrices  were inferred from strategies defined over 1,200 breaths ($\sim$50 minutes) and costs tallied by 24-hour decreases in 4-hour mean P:F ratio as in Application 2.
Figure \ref{fig:costseries} shows the consistent approximation for payoff-derived P:F costs; other costs and context-defining window lengths are presented in SI Fig. \ref{fig:cohortB_app3_cost_evolutions}.
In total, 79 patients contribute a total of 501 strategy--cost pairs to each moving 6-hour window of data considered by independent games. 
Payoff matrices were computed from 3000 Monte Carlo iterations, each using 10K strategy--cost data pairs (from $\sim$125K total). 
Breath costs were derived from the nearest consistent approximation to the median payoff matrix, with entry-wise differences uniformly below 5\%.

Visible from figure \ref{fig:costseries}, breath-type costs vary over time with corresponding changes present in the underlying payoff matrices.
Type \#8's cost was adopted as a fixed reference point to link each time interval's affine space. 
As in previous applications and unlike the verification assumptions, the type cost structure is \emph{not} inversely related to occurrence rates.
The more common types (those with low numbers) tend to have lower costs and therefore less association with injury, although the second most commonly occurring (\#2, featuring low PEEP and driving pressure in stiff lungs) is most strongly tied to a decrease in P:F ratio overall.
The effect may stem from type \#2’s longer breath period (3.8 s \textit{vs.} 2.8 s) or lower driving pressure (11.9 \textit{vs.} 12.6 cm H\textsubscript{2}O).	
Breath type costs vary over time with patient status and therapeutic strategies, and this variability ---across types, time, and chosen consequences--- means VILI minimization may require selecting among multiple time-varying optimal ventilator settings.

\section{Discussion}
This work analyzes data from patients undergoing managed mechanical ventilation (MV) using a game theory-inspired inverse framework in which pairwise comparisons of observed breath types are used to infer their relative consequences. 
The data-generating process (J6) integrates care decisions affecting the patient--ventilator system, enabling the linkage of MV management actions to their downstream effects on patient health status.
In this framework, the joint patient--ventilator--care state space is discretized by relevant features: patient phenotypes \cite{gordon2024icu}, care elements (\textit{e.g.}, ventilator settings), and similarity among MV breath types that include local patient--ventilator interactions detailed by continuous waveform data.
The analysis infers payoff matrices that link breath type occurrence variations to outcome disparities, connecting breath behaviors to observable clinical targets and proxy measurements of ventilator-induced lung injury (VILI).
These payoffs provide a richer consequence representation than direct type-cost estimates and better reflect the comparative decision processes used in clinical practice and reinforcement learning (RL).

Game inversion overcomes barriers to MV data analysis and provides a framework for evaluating observed MV management practices and hypothesized improvements.
It allows direct comparison of patient--ventilator behavior within patient-care regimes (contexts) by quantifying their consequences and provides the relative comparison needed to support RL applications targeting VILI reduction through MV management.
Beyond its role in MV RL problem formulation, the payoff matrices inferred from population game inversion provide ways to address questions such as: \textit{Among the set of comparable strategy--cost pairs in a given context $(\{\pi_n,q_n\}_c)$, is breath type $\phi_i$ associated with more or less injury than $\phi_j$?} and \textit{Which are potentially the best and worse states among $\Phi=\{\phi_k\}$ in the context $c$}?
Clinical data applications demonstrated how game-inferred payoff matrices address these questions: \S\ref{rwd_app1} compares breath type costs and properties within a limited context, \S\ref{rwd_app2} contrasts payoffs in diagnosis-stratified MV patients, and \S\ref{rwd_app3} examines the evolution of MV consequences in a fixed cohort.

The J6 system ---combining the care, patient, and ventilator--- contains complicated, layered interactions and heterogeneity that remain poorly understood.
Within these applications, game inversion on J6 data revealed that MV consequences vary with respect to context, including both those defined by patient cohort and timeframe.
Moreover, by varying the choice of costs and strategy--cost lags, the inferred payoffs surface the poorly understood timescales over which MV consequences and treatment effects intrinsically emerge (SI \ref{SI:rwd_app3_ptypes}).
These results clarify the nature of the time- and context-dependent RL problem and surface clinically important features in need of deeper investigation.
The presented method offers a systematic framework of data analysis to disentangle these effects and validate the hypotheses informing context and breath type definitions.
For example, one may algorithmically search parametric hypotheses defining breath types, context, and cost to identify payoffs with stable, homogeneous consequence attribution.

\subsection{Benefits and Trade-offs of Pairwise Comparison} \label{gains}
By inferring consequences of J6 behaviors comparatively, the game-based framework captures a broader class of breath-type cost relationships than a linearly ordered one. 
The payoff representation better reflects the nature of consequence consideration used in clinical decision-making and reinforcement learning.
Rather than assuming breath types admit a single consistent ordering, it makes this hypothesis empirically testable by comparing the inferred payoff matrix with its nearest consistent approximation. 
A consistent breath type cost ranking can still be recovered as a summary of the inferred payoffs, but it is no longer imposed \textit{a priori} in the consequence linkage. 
These benefits come at the expense of considering all pairwise comparisons, increasing the inference problem from $\mathcal{O}(N)$ to $\mathcal{O}(N^2)$ over $N$ observed strategy--cost pairs.
Another trade-off is the loss of an absolute cost reference: because the payoff contains only pairwise consequence differences, reconstructed linear costs are identifiable only up to an affine translation.
The framework is therefore better suited to characterizing breath-cost structure than absolute breath costs.
		
Cyclic dominance ($\phi_i<\phi_j$, $\phi_j<\phi_k$, $\phi_k<\phi_i$) in games like Rock-Paper-Scissors did not appear in games analyzing clinical J6 data. 
However, inferred payoffs often violated cyclic closure, suggesting that state categories of these data do not support a consistent linear ordering.
There is no \textit{a priori} justification to exclude such possibilities in the motivating RL problem or other downstream uses of relative comparison. 
Intransitive and inconsistent breath-type cost relationships may arise naturally when consequences are more complex than the linear attribution considered here.
Given the complexity of J6, it may be necessary to consider inconsistent cost systems when tying breath type frequency to non-linear costs or when considering multiple consequences simultaneously. 

\subsection{Comparing payoffs in contexts to differentiate treatment groups} \label{HTEs}
Current lung-protective ventilation protocols are not strongly personalized and do not consider heterogeneous treatment effects (HTEs), although experienced providers often individualize care via protocol deviations.
These MV HTEs may manifest MV cost differences arising from variations in patient traits, physiological states, time periods, and care application factors such as provider skill and intervention timing.
Application \S\ref{rwd_app2} demonstrated that MV breath behavior consequences depend strongly on the `ARDS or not' covariate, but this single diagnosis (almost surely) does not fully define treatment-response equivalent endotypes. 
More generally, HTEs require partitioning the J6 state space into appropriate contexts where breath types align with MV cost-equivalence classes. 
Because inferred payoff matrices characterize the breath-type cost structure within each context, they provide a natural basis for comparing contexts and identifying HTEs algorithmically. 
This suggests inverting the proposed game framework over parametric context definitions to identify those yielding statistically stable, variance-minimizing payoff matrices across patient sub-samples. 
Such a process will depend on breath types being sufficiently granular to distinguish cost-equivalence classes across contexts.
Fortunately, Experiment 3 (\ref{sec:verify}) suggests that merging cost-equivalent breath types within a highly granular categorization may be an effective approach.

\subsection{Limitations} \label{sec:limit}
Despite its advantages, game-based inversion has disadvantages beyond the paired-comparison increase in problem size, including reliance on data discretization through contexts, costs, and MV states.
Large, heterogeneous datasets with mixed data types require binning covariates and dimensional reduction of features to yield tractable problems, and these simplifications necessarily result in information loss.
However, breath types and contexts may be reorganized by incorporating additional covariates, but the need to reduce observational data to simplified features will remain essentially unchanged.

The payoff process assumes specific links between breath behavior and costs that warrant review.
The game-based formulation currently assumes linear interaction between breath type occurrence and attributed cost, as no other local association hypotheses are known.
These assumptions could be relaxed by incorporating a different functional form of the payoff process $\mu(\pi_i,\pi_j)$ to introduce threshold effects.
Specific hypotheses for improving breath type-cost mappings are currently unknown but could be developed and explored as extensions of the linear game-like comparison framework.
Also, breath type cost attribution assumed behavior-to-consequence links at specific, fixed delays. 
The true delays between MV breaths and emergent consequences are unknown, but future work should explore and optimize payoff sensitivity to them.

Strategies defined by windowed breath type occurrence distributions omitted sequence order.
Like occurrence thresholds, breath type order may ultimately impact costs but time dependent effects must ultimately be ignored at some level to yield a computationally tractable problem.
Applications of \S\ref{rwd_app1} and \S\ref{rwd_app2} assumed order effects were unimportant within short 5-minute windows for frequently-updated costs and a large patient cohort, respectively.
However, it is unclear whether the $\sim$50-minute strategies considered in Application 3 (\S\ref{rwd_app3}) were reasonable: The 79-patient cohort may be too small to offset order effects, but shorter strategy windows risk collinear of cost (median P:F update time was 45 minutes in cohort B).
The stability of payoff estimation with respect to strategy window sizes should be tested using larger datasets currently being collected \cite{sottile2025developing}.

Finally, payoff inversion assumes that empirical breath categories are at least granular as equivalence classes of cost.
However, increased breath type resolution increases the number of rarely observed breath types, decreasing the effectiveness of data to constraint of the inversion problem.
Applications combined these underrepresented types into a common, cost-heterogeneous group (\#999).
Although this bundling does not affect other payoff entries (per \S\ref{sec:verify}), it limits identification of consequences among those breath categories.
This is a particular concern when rare breath types may be largely consequential to patient outcomes.
Future work could evaluate whether a breath category set supports unique consequence attribution by verifying Monte Carlo-sampled payoff matrices are stable, lack statistically zero off-diagonal entries, and have entrywise unimodal distributions.

\subsection{Future Work and Concluding Remarks}
Inverse game-based analysis bridges the key gaps (VILI linkage and handling heterogeneity) in quantifying MV care effects.
By identifying static cost structures across context-segregated data, it can form a foundation for further data-driven research.
Consideration of non-stationary processes, simulation of cost accumulation over patient trajectories, and modeling counterfactual MV trajectories are a sequence of crucial ongoing developments.

Relaxing stationarity assumptions and comparing treatment types, like ventilation modes rather than only settings, requires further developments such as a comparison of costs between different contexts.
For example, the sequence of payoffs calculated in Application 3 (\S\ref{rwd_app3}) were linked by asserting a constant cost on one breath type across contexts, but this is not likely true.
Defining a payoff function $\mu$ \textit{across} contexts may require reformulating the bilinear payoff system into a non-dimensional, possibly non-linear function (SI \ref{joining_contexts}).
Development of a cross-context payoff would enable continuous and progressive consequence quantification over complete MV patient trajectories, providing the reward representation needed for RL and counterfactual trajectory evaluation.	

The featured J6 data, despite their size and complexity, remain sparse representatives of the patient-MV breath-care space.
Nevertheless, improving MV management hinges on recognizing the diverse relationships between existing practices and patient--ventilator actions captured in the data.
The result of the current work is an inverse game-derived payoff system, a necessary element of reinforcement learning's reward function.
This facilitates the formulation of an RL problem for optimal MV management policies that target the more complete scope of the MV decision-to-consequence process.

RL can now be formulated in narrow contexts to explore action-to-consequence implications and hypothesize cost-minimizing MV management protocols (\textit{i.e.}, policies for action).
SI \ref{sec:RLform} formulates the RL problem over a state-space model using empirical transition probabilities for patient breath types, current actions (care or ventilator changes), and stochastic states (probabilistic type occurrence under given MV settings).
The RL application envisioned ---and now accessible--- relies on the payoff matrix identified through the game-based analysis presented in this work to define incremental costs or rewards for each action or non-action.

\bibliographystyle{siamplain}
\bibliography{lvsegt}
\appendix 
\section{Application 2 and 3 categories and experiment addenda}
Content of this supplement includes information about data and experiment results in applications 2 and 3.
\subsection{Application 2} \label{SI:rwd_app2_ptypes}
\begin{figure}[htb]
    \centering
    \includegraphics[width=.5\textwidth]{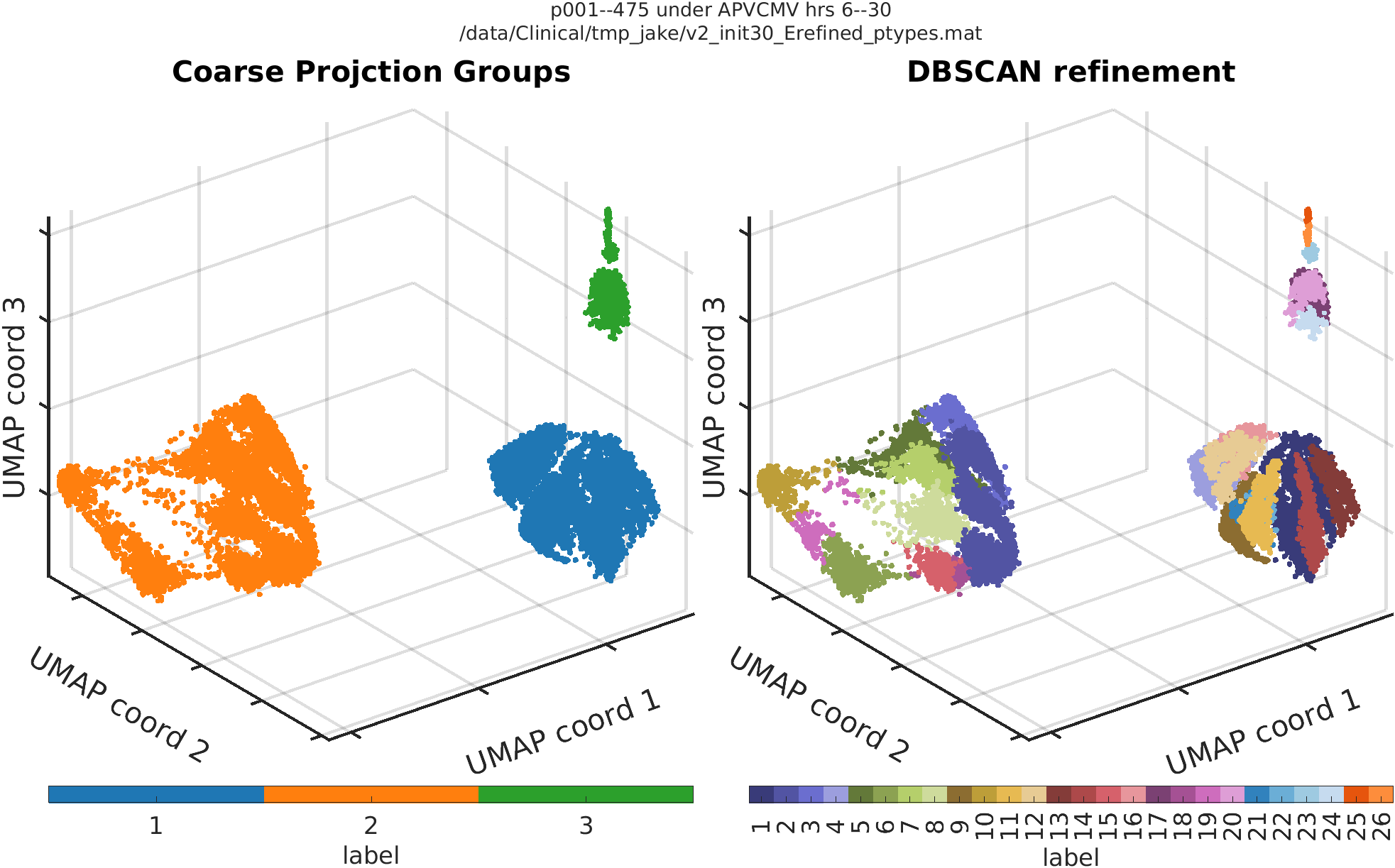}
    \caption{These are the breath types from Application 2 as defined by similarity structure of parametrically represetned waveform statistics (see \cite{stroh2023hypothesis}) and main ventilator settings of hte APVcmv mode.
    At left, UMAP based similarity identifies three globally distinct groups of joint waveform+ventilator settings data. 
    At right, application of Density-based spatial clustering of applications with noise (DBSCAN) to those coarse groups identifies a total of 26 locally similar groups adopted as categories. 
    }
    \label{fig:cohortB_app2_cluster}
\end{figure}

\begin{figure}[htb]
    \centering
    \includegraphics[width=.7\textwidth]{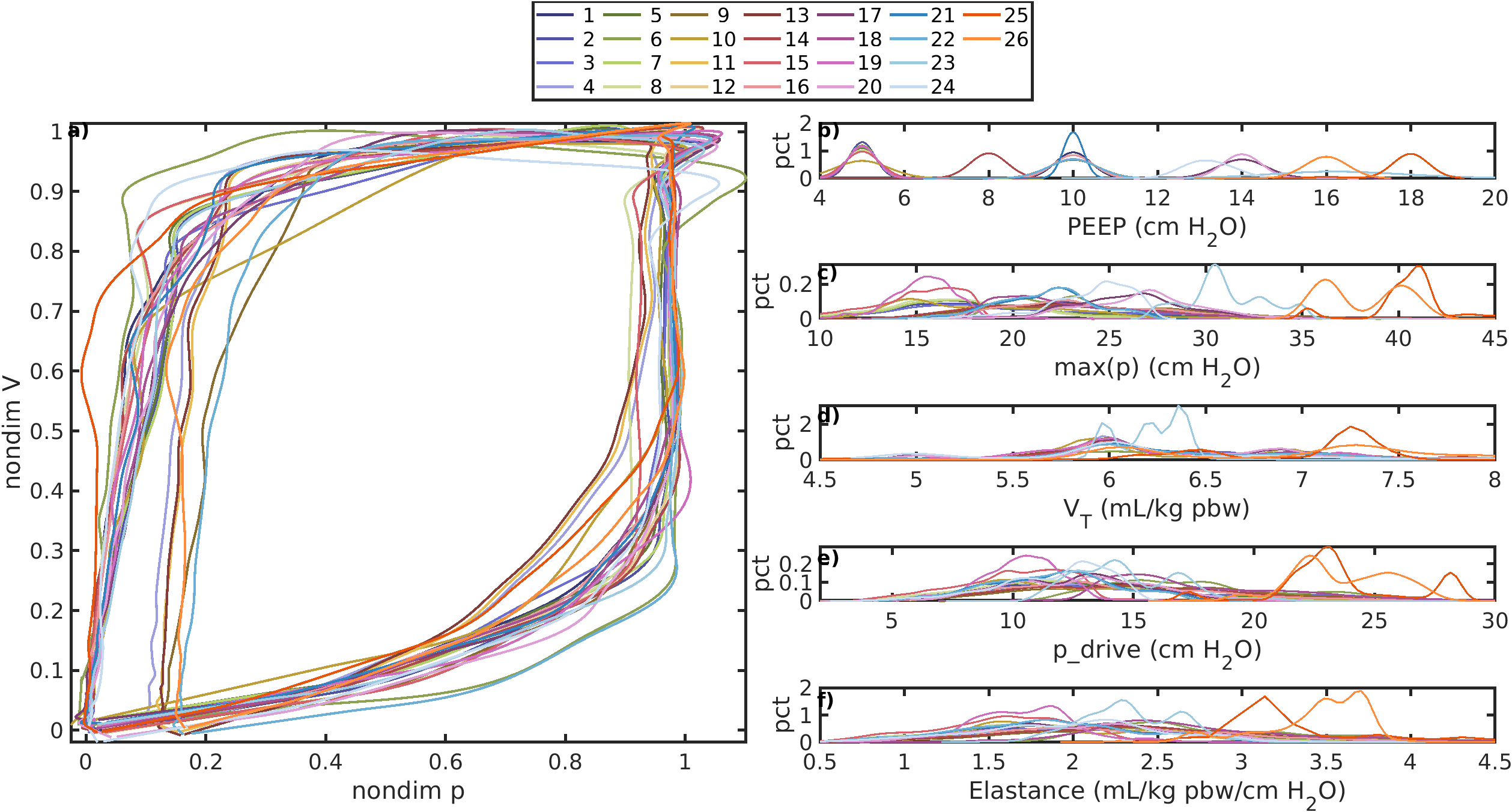}
    \caption{This figure depicts characteristics of the breath categories from Application 2. 
    The 10 most populous categories are retained while remaining types are bundles into a common diverse category in experiments. 
    }
    \label{fig:cohortB_app2_char}
\end{figure}

Application 2 uses the following categories for J6 data, covering the context of the 6--30hr period for breaths under APVcmv while in semi-Fowler's position.
Figures \ref{fig:cohortB_app2_cluster} and \ref{fig:cohortB_app2_char} depict the similarity structure and waveform/settings characterization of the data, respectively.

Table \ref{tab:cohortB_app2_tab} summarizes key features

\begin{figure}[htb]
	\centering
	\includegraphics[width=0.8\textwidth]{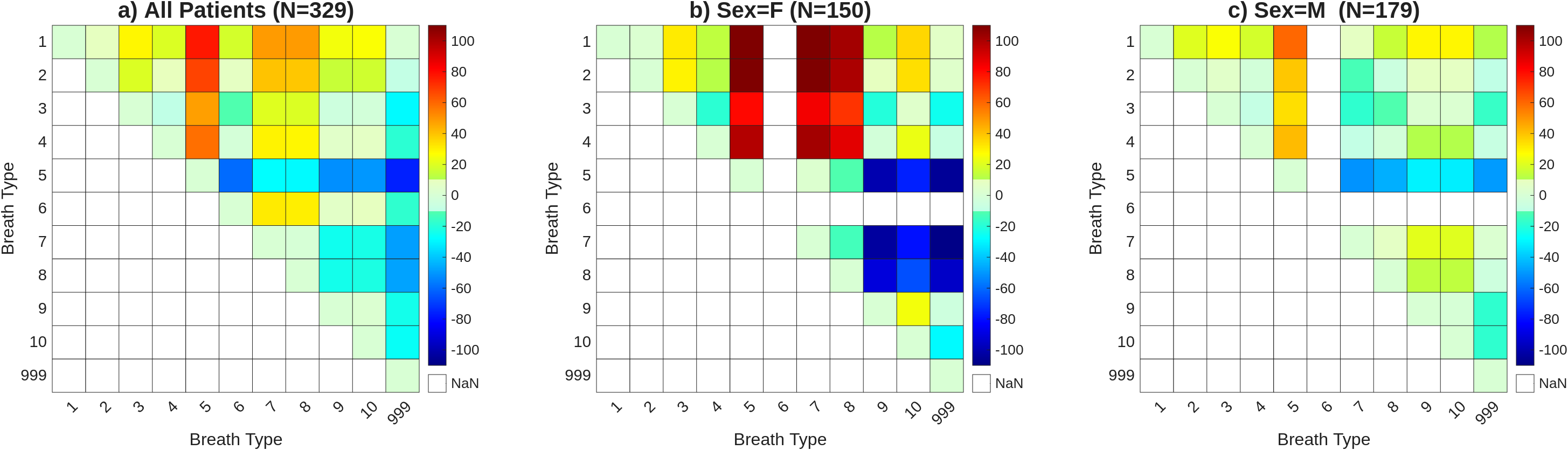}
	\caption{Payoffs for null and an alternate hypothesis in Application 2.
	Heatmaps of the corresponding payoffs for the null hypothesis of an unsegregated population (a), for female patients (b), and for male patients (c).
	The cost structures are overall similar, and differ greatly from those of ARDS-segregated cost structures in the main text.
	Breath type \#6 is not sufficiently resolved in women; for fair comparison, it is bundled into the mixed group (\#999) in the sex-segregated groups.
	The results of Application 2 for ARDS-stratified patients of the same data are evidently not spurious, but rather surface important differences in the consequences of different breath behaviors between patients with and without ARDS.
	}
	\label{fig:null_and_sex_app2}		
\end{figure}

\begin{table}[htb]
    \centering
    \caption{Meta data characterizing breath categories used in Application 2. 
    Headings correspond to: breath type label/id, percentage of total considered data, number of patients represented, breath period, positive end expiratory pressure (PEEP), maximum pressure, tidal volume, and elastance.    
    The second header row identifies units associated with the column.
    Columns 4-though-9 report gaussian parameter of values in `mean(std)' format.
    }
    \label{tab:cohortB_app2_tab}    
    \begin{tabular}{lll lll lll}            
        id & of total & $N_\text{pat}$ & $\theta$  & PEEP       & $p_\text{max}$  & $p_\text{drive}$ & $V_T$      & E\\ 
 & \% &  & s  & cm\,H\textsubscript{2}O & cm\,H\textsubscript{2}O & cm\,H\textsubscript{2}O & mL/kg & cm\,H\textsubscript{2}O$\cdot$kg/mL\\ 
1 & 16.5     & 201        & 2.5(1.0) & 10(2) & 23.3(7.0) &  14.1(6.9) & 6.1(0.9) & 2.3(1.2)\\ 
2 & 15.3     & 225        & 3.0(1.0) & 5(0) & 18.0(6.9) &  12.9(6.7) & 6.3(1.0) & 2.0(1.1)\\ 
3 & 5.7     & 215        & 3.0(1.0) & 5(0) & 19.1(8.7) &  14.0(8.7) & 6.3(1.0) & 2.2(1.3)\\ 
4 & 5.5     & 174        & 2.6(0.8) & 8(2.0) & 22.9(6.2) &  14.1(6.3) & 6.1(0.7) & 2.3(1.0)\\ 
5 & 5.4     & 213        & 3.0(1.3) & 5(0) & 17.6(6.0) &  12.6(5.8) & 6.2(1.0) & 2.0(0.9)\\ 
6 & 4.5     & 156        & 3.0(0.6) & 5(0) & 23.0(5.4) &  17.3(5.7) & 6.3(1.4) & 2.7(1.0)\\ 
7 & 4.4     & 215        & 3.0(0.9) & 5(0) & 17.5(5.4) &  12.4(5.3) & 6.2(1.0) & 1.9(0.9)\\ 
8 & 4.4     & 210        & 3.0(1.2) & 5(0) & 16.9(4.9) &  11.8(4.9) & 6.2(1.0) & 1.8(0.8)\\ 
9 & 3.9     & 158        & 2.5(0.9) & 8(0) & 23.9(8.0) &  15.7(7.2) & 6.1(0.7) & 2.5(1.1)\\ 
10 & 3.5     & 167        & 4.3(0.3) & 5(0) & 17.8(7.1) &  12.8(7.2) & 6.0(1.1) & 2.0(1.0)\\ 
11 & 3.1     & 131        & 2.5(0.7) & 8(0) & 22.6(6.6) &  14.6(6.6) & 6.1(0.8) & 2.3(1.0)\\ 
12 & 3.1     & 132        & 2.6(0.7) & 8(0) & 22.6(6.4) &  14.6(6.4) & 6.1(0.7) & 2.3(1.0)\\ 
13 & 3.1     & 131        & 2.5(0.7) & 8(0) & 22.6(6.4) &  14.6(6.4) & 6.1(0.8) & 2.3(1.0)\\ 
14 & 3.1     & 128        & 2.6(0.8) & 8(0) & 22.6(6.7) &  14.6(6.7) & 6.1(0.9) & 2.3(1.0)\\ 
15 & 2.7     & 183        & 3.2(0.9) & 5(0) & 15.5(3.2) &  10.4(3.3) & 6.2(1.0) & 1.6(0.6)\\ 
16 & 2.3     & 91        & 2.5(1.0) & 10(0) & 23.3(6.3) &  13.2(6.4) & 6.1(0.9) & 2.1(1.2)\\ 
17 & 2.3     & 24        & 2.3(0.7) & 14(2) & 26.1(3.9) &  12.8(4.2) & 6.1(0.9) & 2.0(0.8)\\ 
18 & 2.3     & 207        & 2.7(1.0) & 5(0) & 22.0(5.5) &  16.7(5.6) & 6.2(1.0) & 2.6(0.8)\\ 
19 & 1.8     & 142        & 3.8(0.2) & 5(0) & 15.5(2.2) &  10.5(2.2) & 6.0(0.7) & 1.7(0.4)\\ 
20 & 1.8     & 24        & 2.3(0.7) & 14(0) & 26.9(4.0) &  13.1(4.7) & 6.2(0.9) & 2.1(0.8)\\ 
21 & 1.6     & 84        & 2.6(1.0) & 10(0) & 22.0(3.4) &  11.9(3.7) & 6.3(1.0) & 1.9(0.7)\\ 
22 & 1.6     & 83        & 2.6(1.0) & 10(0) & 21.9(3.3) &  11.8(3.6) & 6.2(1.0) & 1.8(0.7)\\ 
23 & 0.8     & 14        & 2.8(0.7) & 16(0) & 30.7(2.7) &  14.3(3.6) & 6.2(0.4) & 2.3(0.5)\\ 
24 & 0.5     & 23        & 2.3(0.6) & 12(0) & 24.6(3.0) &  12.4(3.3) & 6.0(1.0) & 2.0(0.7)\\ 
25 & 0.5     & 4        & 1.7(0.3) & 18(0) & 41.0(3.9) &  23.2(3.8) & 7.2(1.0) & 3.2(1.0)\\ 
26 & 0.3     & 11        & 2.0(0.4) & 14(0) & 37.6(3.9) &  23.2(3.3) & 7.2(1.3) & 3.5(0.4)        
    \end{tabular}
\end{table}

\subsection{Application 3 Addenda} \label{SI:rwd_app3_ptypes}
\begin{figure}[htb]
    \centering
    \includegraphics[width=.7\textwidth]{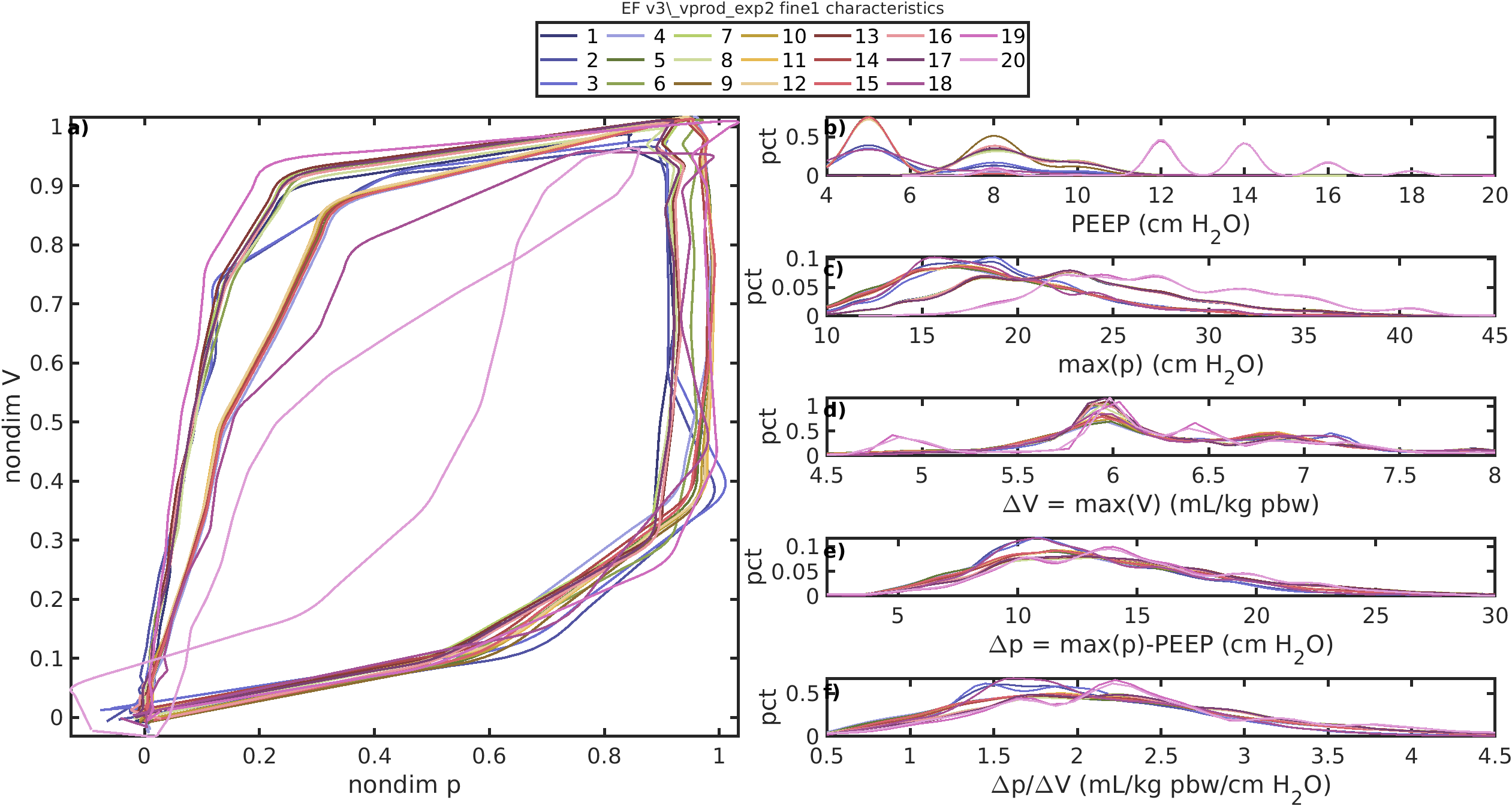}
    \caption{     This figure depicts characteristics of the breath categories from cohort B used to define the strategies in each context (6hr time periods of breaths the first 3 days for 3--7day MV encounters under APVcmv mode and in semi-Fowler's position) Application 3.}

    \label{fig:cohortB_app3_char}
\end{figure}

Application 3 uses breath categories computed from all of cohort B data, although the experiment covers J6 data from the 0--72hr period for breaths under APVcmv while in semi-Fowler's position for 79 patients of cohort B with 3+days of MV and P:F recorded in the 72--96 hour window.
Figures \ref{fig:cohortB_app3_char}  depict waveform/settings characterization of these categories and Table \ref{tab:cohortB_app3_tab} summarizes key features.

\begin{table}[htb]
    \centering
    \caption{These are meta data for breath types in Application 3. The organization follows the previous table.}
    \label{tab:cohortB_app3_tab}    
    \begin{tabular}{lll lll lll}    
        id & of total & $N_\text{pat}$ & $\theta$  & PEEP       & $p_\text{max}$  & $p_\text{drive}$ & $V_T$      & E\\ 
 & \% &  & s  & cmH\textsubscript{2}O & cmH\textsubscript{2}O & cmH\textsubscript{2}O & mL/kg & cmH\textsubscript{2}O$\cdot$kg/mL\\ 
1 & 9.6     & 247        & 2.8(1.0) & 8.0(2.0) & 22.2(7.8) &  13.7(7.1) & 6.1(1.0) & 2.2(1.2)\\ 
2 & 9.1     & 365        & 3.8(1.2) & 5.0(3.0) & 18.1(5.9) &  11.9(5.5) & 6.2(1.1) & 1.9(1.0)\\ 
3 & 5.5     & 359        & 3.8(0.6) & 5.0(3.0) & 18.6(5.9) &  12.1(5.4) & 6.2(1.1) & 1.9(1.0)\\ 
4 & 4.9     & 337        & 3.0(1.0) & 5.0(0.0) & 17.9(6.9) &  12.6(6.7) & 6.3(1.1) & 2.0(1.2)\\ 
5 & 4.8     & 283        & 2.8(1.0) & 5.0(0.0) & 17.8(6.8) &  12.6(6.6) & 6.2(1.1) & 2.0(1.2)\\ 
6 & 4.8     & 243        & 2.8(1.0) & 8.0(2.0) & 22.2(7.8) &  13.7(7.1) & 6.1(1.0) & 2.2(1.2)\\ 
7 & 4.8     & 244        & 2.8(1.0) & 8.0(2.0) & 22.4(7.8) &  13.9(7.1) & 6.1(1.0) & 2.2(1.2)\\ 
8 & 4.7     & 244        & 2.8(1.0) & 8.0(2.0) & 22.3(7.9) &  13.8(7.2) & 6.2(1.1) & 2.2(1.2)\\ 
9 & 4.7     & 274        & 2.8(0.9) & 5.0(0.0) & 18.0(6.7) &  12.8(6.5) & 6.2(1.1) & 2.1(1.1)\\ 
10 & 4.7     & 274        & 2.8(0.9) & 5.0(0.0) & 18.0(6.7) &  12.8(6.5) & 6.2(1.1) & 2.1(1.1)\\ 
11 & 4.7     & 262        & 2.8(1.0) & 5.0(0.0) & 18.1(6.6) &  12.9(6.5) & 6.2(1.0) & 2.1(1.1)\\ 
12 & 4.7     & 262        & 2.8(1.0) & 5.0(0.0) & 18.1(6.7) &  12.9(6.6) & 6.2(1.0) & 2.1(1.1)\\ 
13 & 4.7     & 245        & 2.8(1.0) & 8.0(2.0) & 22.4(7.8) &  13.9(7.1) & 6.1(1.0) & 2.2(1.2)\\ 
14 & 4.7     & 273        & 2.8(1.0) & 5.0(0.0) & 18.1(6.7) &  12.9(6.5) & 6.2(1.0) & 2.1(1.1)\\ 
15 & 4.6     & 264        & 2.8(1.0) & 5.0(0.0) & 17.9(6.6) &  12.7(6.5) & 6.2(1.1) & 2.0(1.1)\\ 
16 & 4.6     & 245        & 2.8(1.0) & 8.0(2.0) & 22.1(7.5) &  13.6(6.9) & 6.1(1.0) & 2.2(1.1)\\ 
17 & 4.5     & 244        & 2.8(1.0) & 8.0(2.0) & 22.2(7.6) &  13.7(6.9) & 6.1(1.0) & 2.2(1.1)\\ 
18 & 3.8     & 334        & 4.3(1.0) & 5.0(0.0) & 18.0(6.0) &  12.2(5.6) & 6.3(1.1) & 1.9(0.9)\\ 
19 & 3.1     & 68        & 2.5(0.9) & 14.0(2.0) & 26.9(8.6) &  14.1(6.7) & 6.2(0.9) & 2.3(1.1)\\ 
20 & 3.1     & 69        & 2.5(0.9) & 14.0(2.0) & 27.0(8.7) &  13.9(6.8) & 6.2(0.9) & 2.2(1.1) 
                        
    \end{tabular}
\end{table}

\begin{figure}[htb]
    \centering
    \includegraphics[width=\textwidth]{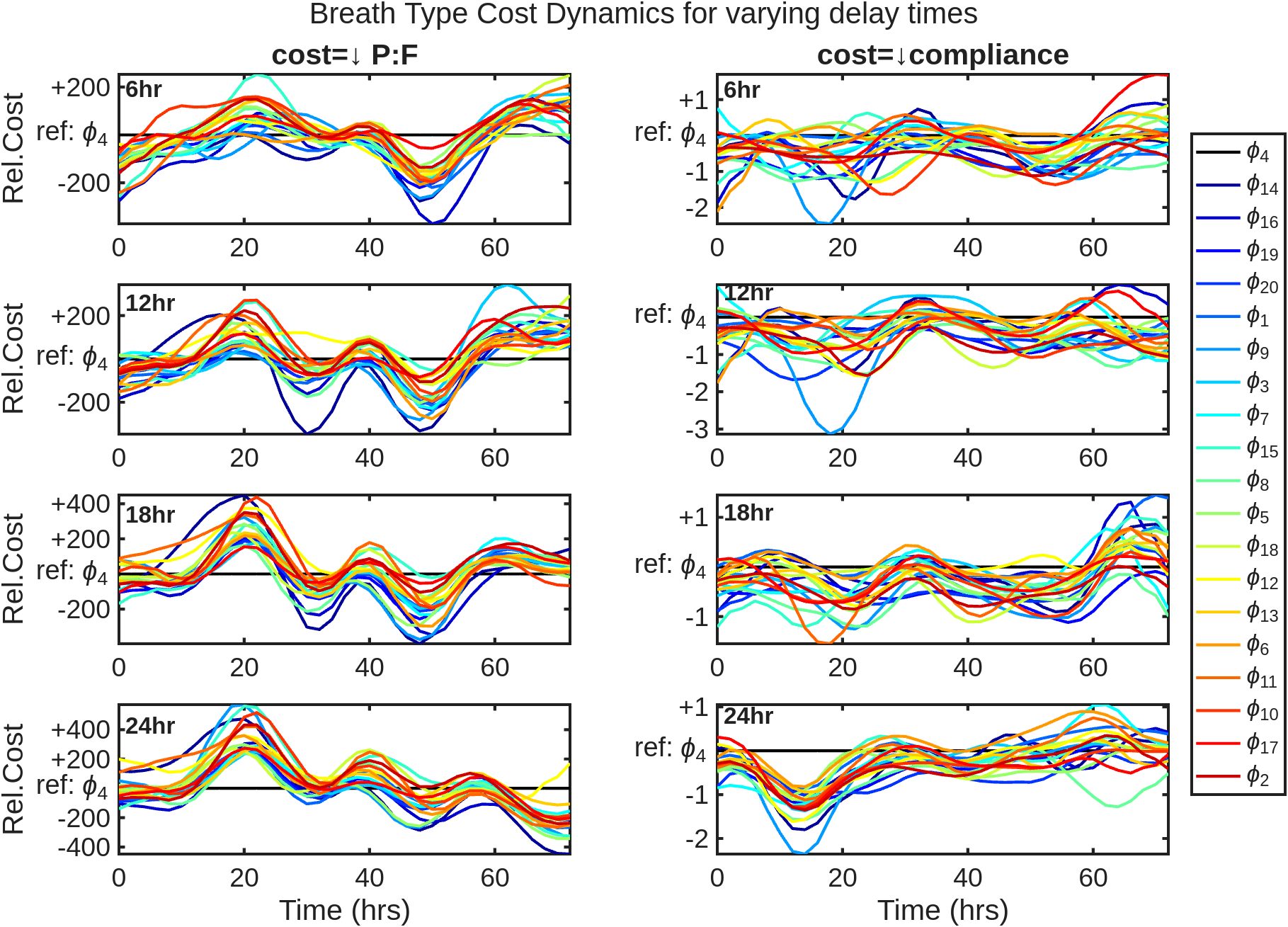}       
    \caption{ Breath-type cost distribution and evolution for different delays between strategy window and associated consequence. 
    Rows show repeated results for experiment 3 using  6:6:24 hour window offsets for costs P:F (\textit{left}) and compliance (\textit{right}).
    Costs appear to have different natural scales of offset due to natural delays between behaviors and consequences in physical processes.
    Identifying the correct offset time for a given cost and consequence are part of the scientific effort motivating this work.
    }
    \label{fig:cohortB_app3_cost_evolutions}
\end{figure}

\section{Unifying context-costs} \label{joining_contexts}
Each game $\Gamma_c$ is formulated according in a particular context $c\in\mathcal{C}$ as $\mu_c(x,y)=x^\top\mathbf{P}_c y=q_x-q_y$ for strategies $x,y\in\Delta^{K-1}$.
The identified matrix $\mathbf{P}_c$ encodes the differences in cost $q$ associated with the breath types $\{\phi^c_i\}_{i=1}^{K_c}$, which act as coordinates for both strategies $x,y$ and $\mathbf{P}$.
Advancing the GT-framework toward include replicator dynamics and Reinforcement Learning requires a cost/outcome function $\mathfrak{u}(x,y)$ across all possible observed states, including the case where $x$ and $y$ are elements of different contexts.
For this development, a general payoff function $\mathfrak{u}$ can be constructed to compare strategies (or individual breath types) from different contexts.
However, the bilinear and affine nature of game payoffs prevents a direct generalization compatible with each $\mu_c$. 
The no information is known about each context's breath-type costs reference values, because only cost differences are inferred but the reference values should be uniform across all contexts.
Linearity in each argument of the payoff process is still desirable to retain additivity so that strategy comparisons can be determined from breath type comparisons.

One formulation of $\mathfrak{u}$, which omits bilinearity while retaining the skew symmetry of $
\mu$, is to compare breath type by the difference in relative quality within their respective games.
For example, if $\phi^c_i$ and $\phi^{c'}_j$ are breath type associated with different contexts ($c$ and $c'$, respectively), they might be compared by differences in regret, or differences relative to the least cost breath type:
\begin{equation}
    \mathfrak{u}(\phi^c_i,\phi^{c'}_j) = \frac{ \text{max}_{k,l\ne j} \mathbf{P}_{c'} (k,l)}{\text{max}\mathbf{P}_{c'}} - \frac{ \text{max}_{k,l\ne i} \mathbf{P}_c(k,l) } { \text{max}{\mathbf{P}_c} }
\end{equation}

Another option is to compare their breath type cost differences reference strategies
\begin{equation}
    \mathfrak{u}(\phi^c_i,\psi^{c'}_j) = \frac{(\phi^c_i)^\top \mathbf{P}_c z_c}{\text{max}\mathbf{P}_c} - \frac{(\phi^{c'}_j)^\top \mathbf{P}_{c'} z_{c'}}{\text{max}\mathbf{P}_{c'}} \nonumber\\
\end{equation}
where $z_c$ is, \textit{e.g.}, the mean $\mathbf{1}_{K_c}/K_c$ (with $K_c$ denoting the number of breath type in context $c$) or an empirical mean of strategies observed in data.
Regardless, such formulations will preserve the sign of $\mu$ in $\mathfrak{u}$ and provide a quantitative comparison when context of compared context categorizations agree.
However, the resulting global payoff function $\mathfrak{u}$ is a separable bilinear form that does not agree structurally with $\mu$ when contexts agree.

\section{Formulating Reinforcement Learning} \label{sec:RLform}
The local cost structure for MV breath types from many example trajectories provides a pathway to Reinforcement Learning (RL) applications.
RL includes Monte Carlo-type methods approximating a solution to a Bellman-like equation. 
The Bellman equation, which quantifies the influence of each decision on the overall outcome, is crucial for identifying optimal strategies, though its direct computation is typically intractable.
Intra-context comparisons ascribe bulk cost information to breath types, while inter-context comparisons supply the decision-cost relationship information necessary for RL.

\begin{figure}[htb]
    \centering
    \resizebox{0.5\textwidth}{!}{
        \begin{tikzpicture}[
          every node/.style={circle, draw, minimum size=1cm, line width=2pt,font=\Large},
          cluster/.style={dashed, draw, ellipse, inner sep=0.1cm},
          thickarrow/.style={-{Latex[length=3mm]}, line width=2.5pt},
          thinarrow/.style={<->, shorten >=2pt, shorten <=2pt,line width=1.25pt}
        ]

        \node (c1s1) at (0.5,2) {$\phi_1^1$};
        \node (c1s2) at (2.5,2) {$\phi_2^1$};
        \node (c1s3) at (0.5,0) {$\phi_3^1$};
        \node (c1s4) at (2.5,0) {$\phi_4^1$};

        \node[cluster, fit=(c1s1) (c1s2) (c1s3) (c1s4), label=above:$c_1$,font=\Large] (cluster1) {};

        \draw[thinarrow,red, line width=1.5pt] (c1s1)--(c1s2);
        \draw[thinarrow] (c1s1)--(c1s3);
        \draw[thinarrow,red, line width=1.5pt] (c1s1)--(c1s4);
        \draw[thinarrow,red, line width=1.5pt] (c1s2)--(c1s3);
        \draw[thinarrow] (c1s2)--(c1s4);
        \draw[thinarrow,red, line width=1.5pt] (c1s3)--(c1s4);

        \node (c2s1) at (7.5,2) {$\phi_1^2$};
        \node (c2s2) at (9.5,2) {$\phi_2^2$};
        \node (c2s3) at (7.5,0) {$\phi_3^2$};
        \node (c2s4) at (9.5,0) {$\phi_4^2$};
        \node[cluster, fit=(c2s1) (c2s2) (c2s3) (c2s4), label=above:$c_2$] (cluster2) {};


        \draw[thinarrow] (c2s1)--(c2s2);
        \draw[thinarrow,red, line width=1.5pt] (c2s1)--(c2s3);
        \draw[thinarrow,red, line width=1.5pt] (c2s1)--(c2s4);
        \draw[thinarrow,red, line width=1.5pt] (c2s2)--(c2s3);
        \draw[thinarrow,red, line width=1.5pt] (c2s2)--(c2s4);
        \draw[thinarrow] (c2s3)--(c2s4);

        \node (c3s1) at (4,-4) {$\phi_1^3$};
        \node (c3s2) at (6,-4) {$\phi_2^3$};
        \node (c3s3) at (4,-6) {$\phi_3^3$};
        \node (c3s4) at (6,-6) {$\phi_4^3$};
        \node[cluster, fit=(c3s1) (c3s2) (c3s3) (c3s4), label=above:$c_3$] (cluster3) {};

        \foreach \i/\j in {c3s1/c3s2, c3s1/c3s3, c3s1/c3s4, c3s2/c3s3, c3s2/c3s4, c3s3/c3s4} {
          \draw[thinarrow] (\i) -- (\j);
        }

        \draw[thickarrow, blue, <->] (cluster1.east) -- (cluster2.west);
        \draw[thickarrow, blue,<->] (cluster1.south) -- (cluster3.north west);
        \draw[thickarrow, blue,<->] (cluster2.south) -- (cluster3.north east);

        \end{tikzpicture}
    }
    \caption{Stochastic breath type transition model underlying the RL formulation, with arrows indicating transitions between breath type. 
    Dashed lines outline three distinct contexts that are mutually accessible by changes in care (\textit{e.g.}, vent mode, position, sedation, or other drug administration such as vasopressor use) or patient phenotype (\textit{e.g.}, gross health status).
    Bidirectional changes (blue) between contexts are not universal (\textit{e.g.}, time-changing contexts are directed) as illustrated here.
    Within contexts, the limited ventilator settings (PEEP, V\textsubscript{T}, \textit{et al.}) can be altered by care processes (red).
    However, the context and vent settings do determine the system, as waveform components vary under those conditions (black).
    Integrated transitions over short times define strategies, the states defined for RL below.}    
    \label{fig:statespacemodel}
\end{figure}

Figure \ref{fig:statespacemodel} depicts a finite state space model over three interchangable contexts to illustrate the RL application formulated below, with colors depicting example MV state changes influenced by actions (red, blue) and uncontrolled dynamics (black) corresponding to the patient.
This model differs greatly from the model formulated in a body of MV-RL work -largely on MIMIC-sourced data- where states are defined by hour-scale averages of lab measurements, patient demographics, and medication/fluid input/output \cite{peine2021development,kondrup2023towards,liu2024reinforcement}. 
This formulation is more related to works that include lung physiologically-related parameters and measurements such as plateau pressure \cite{yu2020supervised} or an array of pressure parameters ( mean, plateau, drive), compliance, and dead space volume as well as posture and other ventilator details \cite{roggeveen2024reinforcement}. 
States incorporate distributional data that define local breath types in each context, including representations of observed ventilator waveforms.

\begin{enumerate}
\item Let $\mathcal{S}$ be the \textit{state space} of the whole game system, so $\mathcal{S}=\oplus_{c\in\mathcal{C}} \Delta_c^{K_c-1}$ meaning that the state space comprises all possible strategies in all possible contexts. 

\item Each state $\pi \in \mathcal{S}$ is a strategy vector of coefficients $\pi = (\phi_1, \phi_2, \ldots, \phi_\kappa)$ corresponding to the full range of categories $\{\phi_k\}_1^\kappa$. 
This way, each context defines an independent subspace of behaviors in $\mathcal{S}$ while still associating directly with $\pi \in \Delta_c^{K_c-1}$ in its native context by projection.

\item Each breath type $\phi_k$ is comprised of three components $(w_k, v_k, c_k)$ where $w$ describes the a waveform shape (patient-patient ventilator interaction), $v$ describes the ventilator settings (local MV care aspects), and $c$ describes the native context (patient phenotype, care modes) associated with the breath type.
    
\item The \textit{action space} $\mathcal{A}$ consists of context-settings pairs $(v, c)$ that can be changed and manipulated by care processes.

\item The \textit{states} are sampled from distributions and are subject to stochastic dynamics described probabilistically with $w(t) \sim p(w\,\mid\,v, c)$.

\item The transition probability distribution $T_\mathcal{S}(\pi',\pi | v,c) = p(\pi'\,\mid\, \pi, (w|v,c) )$ over the state transitions $\pi\to\pi'$ describes the transition probability of starting at state $\pi$, selecting action $(v,c)$, and observing some distribution on observed waveforms $w|v,c$ to define $\pi'$.

Note that strategies (\textit{i.e.}, windowed breath type distribution) $\phi$ is co-determined by stochastic  $w$ and action-specified $(v,c)$.
\textit{Actions (changing modes or other external details of care and vent settings) change constraints of the patient-ventilator behavior, but do not fully determine it.}

\item The \textit{reward/cost function} $R(\pi, (v, c), \pi')$  is the generalized breath type comparison function $\mathfrak{u}(\pi,\pi')$ described in \S\ref{joining_contexts}.
The reward function may also be defined from the EGT payoff function when actions do not alter context.
Note that the actions hope to maximize reward/minimize cost of the stochastic dynamics (\textit{i.e.}, the probability distribution of breath type).

\item The function $\varpi: \mathcal{S} \to \mathcal{A}$ is a \textit{policy} determining how actions should be taken based on system state.
\end{enumerate}

These terms/definitions formulate the trajectory value function $V$ under policy $\varpi$ as
\begin{equation}
    V^{\varpi}(\pi) = \mathbb{E}\left[ \sum_{t=0}^{\infty} \gamma^t R(\pi_t, \varpi(\pi_t), \pi_{t+1}) \right]
\end{equation}
where $\gamma \in [0,1]$ is a parameter associated with the discount rate on state-based rewards over time.
This expectation is over $\{\pi_t\}_{t\ge0}$ with the following chain of dependence:
\begin{enumerate}
    \item $(v_t,c_t) = \varpi(\pi_t)$ (policy sets action)    
    \item $w_t \sim p(w_t\,|\,v_t, c_t)$ (given action, determine observed breath type distribution)
    \item $\pi_{t+1} \sim p( \pi_{t+1}\,|\,\pi_t,  w_t )$ (dynamic transitions of strategies)    
\end{enumerate}

The research objective targeted in this work is to identify MV policies $\varpi$ that reduce or minimize the total trajectory value (the cost of MV, as measured by VILI-associated measures) for MV patient trajectories.

\section*{Declarations}

\textbf{Contributions}: All authors have made substantial intellectual contributions to the study conception, execution, and design of the work. All authors have read and approved the final manuscript.  In addition, the following contributions occurred:  Conceptualization: DJA, JdW, BJS, JNS; Methodology: DJA, JdW, JNS; Formal analysis and investigation: DJA, JNS; Writing - original draft preparation: DJA, JNS; Writing - review and editing: all; Funding acquisition: DJA, TDB, JdW, GH, BJS, PDS; Resources: DJA, TDB, BJS, PDS; Supervision: DJA, BJS

\textbf{Artificial Intelligence}: The cyborg and practitioner image elements of \cref{fig:cz} were generated from text prompts using chatGPT on August 2, 2025, which were then manually edited and assembled. No other LLM or GPT output appears in this work.

\textbf{Disclosures}: The authors report no conflicts of interest.

\textbf{Data Accessibility}: The clinical datasets used and/or analyzed during the current study are not publicly available due to ongoing collection, lack of patient consent for broad dissemination of their data, and data size (25 GB for cohort A, more than 2TB for cohort B). 
Data access can be initiated by reasonable request to the corresponding author (J.N. Stroh, jn.stroh@cuanschutz.edu) and will require institutional data use agreement.

\end{document}